\documentclass[12pt,preprint]{aastex}

\usepackage{graphicx}
\usepackage{afterpage}

\shorttitle{Formation of planetesimals}
\shortauthors{Michikoshi et al.}

\begin{document}
\title{ 
$N$-body Simulation of Planetesimal Formation Through Gravitational Instability of a Dust Layer
}
 
\author{Shugo Michikoshi\altaffilmark{1}, 
Shu-ichiro Inutsuka\altaffilmark{1}, Eiichiro Kokubo\altaffilmark{2}, and Izumi Furuya\altaffilmark{3}
}
\altaffiltext{1}{
Department of Physics, Kyoto University, Kyoto 606-8502, Japan}
\altaffiltext{2}{
Division of Theoretical Astronomy, National Astronomical Observatory of Japan, Osawa, Mitaka, Tokyo 181-8588, Japan
}
\altaffiltext{3}{
NS Solutions Corporation, 20-15, Shinkawa 2-chome, Chuo-ku, Tokyo 104-8280, Japan
}
\email{shugo@tap.scphys.kyoto-u.ac.jp, inutsuka@tap.scphys.kyoto-u.ac.jp, kokubo@th.nao.ac.jp, and furuya.izumi@ns-sol.co.jp}
\begin{abstract}
We performed N-body simulations of a dust layer without a gas component and examined the formation process of planetesimals. 
We found that the formation process of planetesimals can be divided into three stages: 
the formation of non-axisymmetric wake-like structures, the creation of aggregates, and the collisional growth of the aggregates.  
Finally, a few large aggregates and many small aggregates are formed.  
The mass of the largest aggregate is larger than the mass predicted by the linear perturbation theory. 
We examined the dependence of system parameters on the planetesimal formation. 
We found that the mass of the largest aggregates increase as the size of the computational domain increases. 
However the ratio of the aggregate mass to the total mass $M_\mathrm{aggr}/M_\mathrm{total}$ is almost constant $0.8-0.9$.
The mass of the largest aggregate increases with the optical depth and the Hill radius of particles.
\end{abstract}
\section{Introduction \label{sec:intro}}
Many Jupiter-mass planets have been found during recent observations of extrasolar planets.
The existence of extrasolar planets means that the formation of Jupiter-mass planets is a common process.
According to the standard model of planet formation, the growth of solid planets occurs in two sequential phases.
In the first phase, micron-sized dust grains grow into kilometer-sized bodies. 
These kilometer-sized bodies, called planetesimals, are the building blocks of terrestrial planets and cores of giant planets. In the second stage, the planetesimals continue to grow through inelastic collisions.

The initial stage of coagulation of dust grains is supposed to continue up to about centimeter size.
However, the agglomerative growth of dust grains, from cm to km size, is not well understood. 
The inward orbital drift of meter-sized bodies due to gas drag is very rapid. 
They would fall onto the central star from 1AU within a timescale of only $10^2$ years
\citep{Adachi1976,Weidenschilling1977}.
Thus, the direct assemblage of dust grains in a laminar flow condition seems difficult.
 
One of the scenarios for this stage is the gravitational instability (GI) scenario. 
If particles settle into a layer having sufficiently high density and low velocity dispersion, the dust layer rapidly collapses owing to its self-gravity, forming km-sized planetesimals \citep{Safronov1969,Goldreich1973,Coradini1981,Sekiya1983,Yamoto2004}. 
The timescale of this collapse is only on the order of a Kepler period in this scenario, and the rapid migration of meter-sized bodies can be avoided.

However, there is a critical issue in this GI scenario. As dust grains settle toward the midplane, the rotational velocity around the midplane increases because of the reduced effect of the gas pressure compared to the centrifugal force and the solar gravity. The rotational velocity is a function of vertical distance from the midplane, and the resultant shear may induce Kelvin-Helmholtz instability. The slightest amount of turbulence in the nebular gas due to Kelvin-Helmholtz instability prevents dust grains from settling.
Many authors have investigated this issue \citep{Weidenschilling1980,Cuzzi1993,Champney1995,Weidenschilling1995,Sekiya1998,Dobrovolskis1999,Sekiya2000,Sekiya2001,Ishitsu2002,Ishitsu2003,Michikoshi2006}. 
In the case of a minimum mass solar nebula \citep{Hayashi1981}, they concluded that gravitational fragmentation of the dust layer is difficult if turbulence exists. 

Many authors continue to consider various effects of the GI scenario.
\citet{Sekiya1998} showed that large values of the total dust-to-gas mass ratio may provide a density cusp at the midplane.
\citet{Youdin2002} and \citet{Youdin2004} argued that this cusp can be a trigger of GI.
\citet{Johansen2006a} performed the numerical simulations of the Kelvin Helmholtz instability. 
In the saturated turbulence they found a highly overdense clumps of dust particles related to the streaming instability found by \citet{Youdin2005}.
\citet{Michikoshi2006} performed two fluid analyses of Kelvin-Helmholtz instability in the dusty gas layer.
They confirmed the previous result, that the Kelvin-Helmholtz turbulence must occur before GI. 
However, they proposed a possible path toward planetesimal formation through GI, if the dust could grow into a sphere of about 5$\mathrm{m}$ .
Although studies have been conducted on the condition of GI or the linear regime, 
there has been little study of the non-linear regime of GI.
The purpose of our study is to examine the formation process of planetesimals through using N-body simulations.

From the linear stability analysis of a two-dimensional self-gravitating disk, the condition for gravitational instability of the dust layer is given in terms of Toomre's $Q$ value as \citep{Toomre1964}
\begin{equation}
Q=\frac{\sigma_x \Omega_0}{\pi G \Sigma_\mathrm{d}} < 1,
\end{equation}
where $\sigma_x$ is the radial velocity dispersion, $\Omega_0$ is the Keplerian angular velocity, $G$ is gravitational constant, and $\Sigma_\mathrm{d}$ is the surface density of the dust layer.
The most unstable wavelength $\lambda_{\mathrm{most}}$ of the axisymmetric instability mode for $Q=1$ is given by 
\begin{equation}
\lambda_\mathrm{most}=\frac{2\pi^2 G \Sigma_\mathrm{d}}{\Omega_0^2}.
\end{equation}
If a dust layer becomes unstable, $Q<1$, it breaks up into fragments. The mass of a planetesimal is estimated as 
\begin{equation}
M_{\mathrm{theor}} = \pi \Sigma_\mathrm{d} \left( \frac{\lambda_{\mathrm{most}}}{2}\right)^2.
\end{equation}

At radius $a=1\mathrm{AU}$ of the minimum-mass solar nebula model, the surface density of dust is $\Sigma_\mathrm{d} \simeq 10 \mathrm{g\,cm}^{-2}$. 
\citet{Nakagawa1986} assumed that the gas is laminar during the settling phase and examined the settling path and growth of dust particles due to sweeping process. They found that the size of dust particles reaches 20cm.
The dynamical optical depth is not negligible at $1\mathrm{AU}$ (\S \ref{sec:model}).
The direct collision between particles plays an important role in the evolution of the dust layer.
The dynamical behavior of a self-gravitating collisional particle system has been studied in the context of the planetary ring. 

In earlier studies of planetary rings using N-body simulations, gravitational force was neglected
\citep{Brahic1977}, or only the self-gravity was considered \citep{Wisdom1988}. 
\citet{Salo1995} extended simulations including mutual gravitational force and performed local N-body simulations with various ring parameters. He showed that the wakes are created if both self-gravity and inelastic collisions are included.
In addition, he found that large surface density $\Sigma$ and small restitution coefficient $\epsilon$ lead to gravitational instability and formation of non-axisymmetric wake-like structures and aggregates.
In the case of planetesimal formation, these aggregates might correspond to planetesimals.
\citet{Tanga2004} performed N-body simulations of a gravitationally unstable particle disk in the outer solar system.
They examined the formation of aggregates through gravitational instability.
In the case of the particle disk without gas, they confirmed that the wake-like structures are formed. In addition, they performed the simulation including the effect of gas drag. 
The dissipation due to gas drag changed the nonlinear process of GI. They found the formation of large virialized cluster. 
In their simulation, the optical depth is very small, and the particle disk is unstable, even initially.
If we consider the formation of planetesimals through gravitational instability at $a_0 \simeq 1\mathrm{AU}$, 
the dynamic is collision-dominated. In addition, the initial velocity dispersion must be large owing to turbulent flow.
The particle disk is gravitationally stable in the initial condition of our simulations.

\citet{Furuya2004} performed a set of simulations in the case of large optical depth. 
She adopted the hard-sphere model as a collision model, and examined parameter dependences on the size of computational domain and restitution coefficient. 
She found that the formation process of planetesimal is universal to parameters in the case where initial Toomre's $Q_\mathrm{init}>2$. The formation process is divided into three stages: the formation of non-axisymmetric wake-like structures, the creation of aggregates, and the collisional growth of the aggregates. 

In this study, we perform the high-resolution local N-body simulations and study the gravitational instability of a dust layer. 
We extend Furuya's simulations by changing various parameters and collision models, and examine the formation process of planetesimals. 
We adopt the hard-sphere and soft-sphere models as a collision model. 
In the hard-sphere model, penetration between particles is prohibited. 
The velocity changes in an instant when a collision occurs.
In the soft-sphere model, particles can overlap when they are in contact, and the collision has duration. We study the dependences on the size of computational domain, the Hill radius of the particle, the optical depth, restitution coefficients, and the duration of collision.

In \S \ref{sec:method}, the simulation method, disk model, and initial conditions are described.
Our results are presented in \S \ref{sec:res}.
We explain the formation process of planetesimals through gravitational instability and examine the time evolution of the typical model and its dependence on the disk parameters.
In \S \ref{sec:sum}, we summarize the results.

\section{Methods of Calculation \label{sec:method}}
\subsection{Equation of Motion}
The effect of gas is important if we consider centimeter-sized dust grains. 
However, it is difficult to consider dust particles and gas consistently.
As the first step of our study, we neglect the effect of gas for the sake of simplicity.
In this paper we examine the nature of gravitational instability of a particle disk.
This situation corresponds to the disk of large particles which are not affected by gas.
We will study the effect of gas in the near future.

Suppose a central star with mass $M_*$ and dust grains with mass $m_i$.
We introduce rotating Cartesian coordinates $(x,y,z)$ called \textit{Hill coordinates},
the $x$-axis pointing radially outward, the $y$-axis pointing in the direction of rotation, and the $z$-axis normal to the equatorial plane.
The origin of coordinates moves on a circular orbit with the semi-major axis $a_0$ and the Keplerian angular velocity $\Omega_0$. 
We assume $m_j \ll M$, $|x_j|, |z_j| \ll a_0$, $|\dot x_j|, |\dot z_j| \ll a_0 \Omega_0$, and $|\ddot x_j|, |\ddot z_j| \ll a_0 \Omega_0^2$, where dot denotes time derivatives.
We can write the equation of motion in non-dimensional forms independent of mass and the heliocentric distance $a_0$ if we scale the time by $\Omega_0^{-1}$, the length by Hill radius $h a_0 = r_\mathrm{H}$, and the mass by $h^3 M_*$, where $h=(2 m_\mathrm{p}/3M_*)^{1/3}$, $m_\mathrm{p}$ is the characteristic mass of particles \citep{Nakazawa1988}.
We obtain Hill's equation:
\begin{equation}
\begin{array}{ccccccc}
\displaystyle{\frac{d^2 x_i}{dt^2}} & = &  \displaystyle{2\frac{dy_i}{dt}} &+&3 x_i & + & \displaystyle{\sum_{j=1,j\ne i}^N \frac{m_j}{r_{ij}^3}(x_j-x_i)}, \\
\displaystyle{\frac{d^2 y_i}{dt^2}} & = & \displaystyle{-2\frac{dx_i}{dt}} & &      & + & \displaystyle{\sum_{j=1,j\ne i}^N \frac{m_j}{r_{ij}^3}(y_j-y_i)}, \\
\displaystyle{\frac{d^2 z_i}{dt^2}} & = &                   &-&z_i   & + & \displaystyle{\sum_{j=1,j\ne i}^N \frac{m_j}{r_{ij}^3}(z_j-z_i)}, \\
\end{array}
\label{eq:EOM}
\end{equation}
where $r_{ij}$ is the distance between particles $i$ and $j$.
The first terms on the right-hand side are the Coriolis force. The second and third terms denote the tidal force and mutual gravity between particles, respectively.

The equation of motion is invariant under the transformation $(x,y,z,v_x,v_y,v_z) \to (x+n L_x, y - (3/2)n L_y \Omega_0 t + m L_y , z,v_x,v_y-\frac{3}{2}\Omega_0 n L_x,v_z)$, where $m$ and $n$ are integers, and $L_x$ and $L_y$ are the
size of the computational domain \citep{Wisdom1988}. 
The dynamics are independent of the choice of origin. The motion of particles is pursued only in the cell with periodic boundary conditions. 
We use a sliding cell method in which the image comes into the cell from the opposite boundary when a particle crosses 
the boundary of the cell.

\subsection{Gravity Calculation}
In the original Wisdom and Tremaine's method, the mutual gravity was ignored or was only treated as the vertical force by enhancement of a vertical frequency. 
In order to include the mutual force exactly, we adopt the subregion method introduced by \citet{Daisaka1999}.
Their method has a cut-off of the gravitational force. \citet{Salo1995} studied the influence of the cut-off length. 
A larger cut-off length is needed when the aggregates are formed or the wake-like structures appear. 
The cut-off length in this method is similar to the length of a simulation region that is sufficiently large to create the wake-like structures or aggregates.

\begin{figure}
\plotone{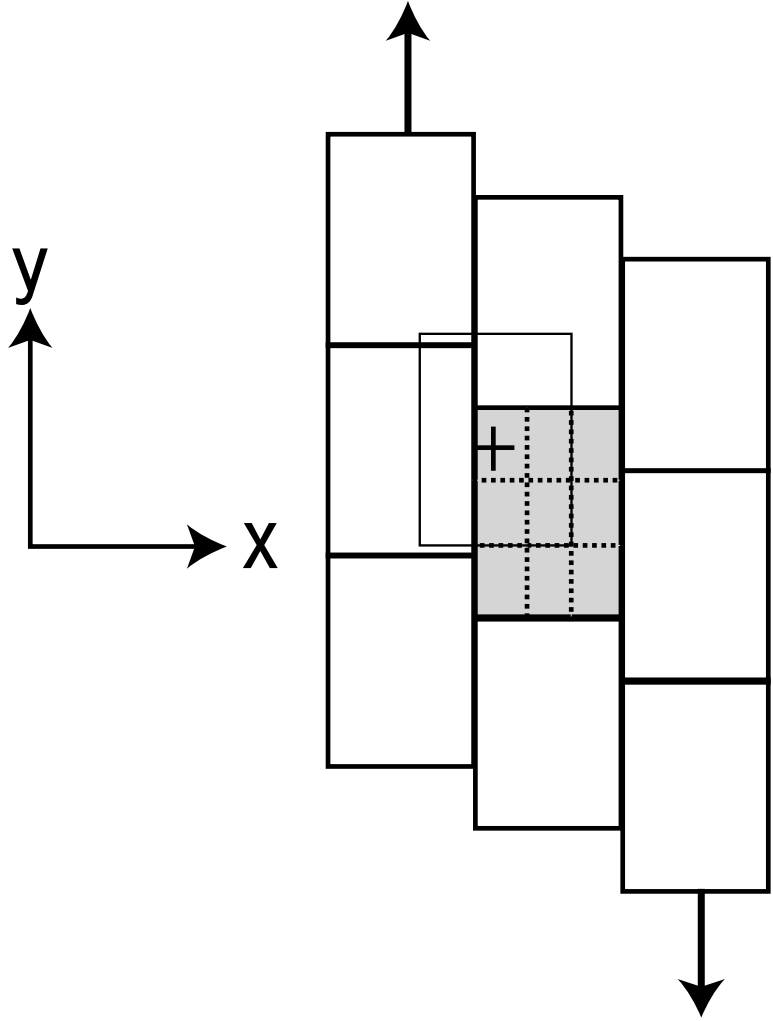}
\caption{Schematic illustration of a computational cell (shaded area surrounded by a thick line) with nine subregions (divided by dotted line) for calculation of the mutual gravitational force. 
The region surrounded by a thin line corresponds to the virtual region for a subregion, which is created to center the subregion and has the same size as a computational cell. }
\label{fig:subregion.eps} 
\end{figure}

We assume an active region and its copies (see Figure \ref{fig:subregion.eps}). 
Inner and outer cells have different angular velocities. These cells slide upward and downward with a shear velocity $3 \Omega_0 L_x/2$.
We divide the simulation region into nine subregions. 
For each subregion, we make virtual regions the same size as the original region. The subregion is centered in the virtual region. For particles in the subregion, we calculate gravitational forces from all particles in its virtual region that contains copied particles.

The most expensive part of calculation is the calculation of gravitational forces whose cost is $O(N^2)$.
We calculate the gravitational force using the special-purpose GRAPE-6 computer \citep{Makino2003}.
With GRAPE, calculation time is significantly reduced.

\subsection{Collision Model}
\subsubsection{Hard-Sphere Model}
A hard-sphere model has been used in previous simulations of planetary rings \citep{Wisdom1988,Salo1991,Richardson1994,Daisaka1999,Daisaka2001}.
In the hard-sphere model, particles are considered to be hard and penetration between particles is prohibited. 
When a collision event occurs, the particle velocity changes in an instant.
In our simulation, a collision conserves the relative tangential velocity and reduces the absolute value of the relative normal velocity by a factor of $\epsilon$.
The relative velocity after a collision is described by
\begin{equation}
\mathbf{v_{ij}'}=\mathbf{v_{ij}}-(1+\epsilon)(\mathbf{n_{ij}}\cdot \mathbf{v_{ij}})\mathbf{n_{ij}},
\end{equation}
where $\mathbf{v_{ij}}$ is the relative velocity and $\mathbf n_{ij}$ is a unit vector along $\mathbf{r_{ij}}$. The conservation of momentum yields
\begin{equation}
m_i \mathbf{v_{i}'}+m_j \mathbf{v_{j}'}=m_i \mathbf{v_{i}}+m_j \mathbf{v_{j}},
\end{equation}
where $\mathbf{v_{i}'}$ and $\mathbf{v_{j}'}$ are velocities of the particle $i$ and $j$ after the collision.
Thus the velocities after the collision are described by
\begin{equation}
\mathbf{v_i'}=\mathbf{v_i}-\frac{m_j}{m_i+m_j}(1+\epsilon)(\mathbf n_{ij}\cdot \mathbf v_{ij})  \mathbf n_{ij},
\label{eq:vv1}
\end{equation}
\begin{equation}
\mathbf{v_j'}=\mathbf{v_j}+\frac{m_i}{m_i+m_j}(1+\epsilon)(\mathbf n_{ij}\cdot \mathbf v_{ij})  \mathbf n_{ij}.
\label{eq:vv2}
\end{equation}
If a pair overlaps, $r_{ij} < r_i + r_j$, and is approaching $\mathbf n_{ij} \cdot \mathbf v_{ij}<0$, this pair collides, and change their velocities by using Equations (\ref{eq:vv1}) and (\ref{eq:vv2}).

\subsubsection{Soft-Sphere Model}
In the soft-sphere model, particles are permitted to overlap when they are in contact \citep{Salo1995}. 
This model allows us to easily calculate the simultaneous collisions of many particles.
The interaction forces are composed of a restoring harmonic force and a viscous damping force,
\begin{equation}
F(\alpha) = k \alpha + \beta \dot \alpha,
\end{equation}
where $\alpha$ is the penetration depth, $k$ is the stiffness of the string, and $\beta$ is the coefficient of viscosity.
The force acts in the direction joining particle centers.
Assuming that there are only two particles, and neglecting the gravitational force, the motions of the pair correspond to the damped oscillation whose frequency is given as $\omega=\sqrt{\omega_0^2 - 1/(2s)^2}$, where $s=\bar m /\beta$, $\bar m$ is the reduced mass, and $\omega_0 = \sqrt{k/\bar m}$ \citep{Salo1995}.
The coefficient of restitution is $\epsilon=\exp\left(- \pi/(2 \omega s)\right)$,
and the duration of the impact is $T_s=2\pi/\omega$ \citep{Dilley1993}.

\subsection{Timestep}
The equation of motion is integrated with a leap-frog scheme.
We adopt the variable timestep used by \citet{Daisaka1999}.
In the case of the hard-sphere model, the formula of timestep is simply given by
\begin{equation}
\Delta t=\eta_1 \min_i \frac{|\mathbf{a_i}|}{|\mathbf{\dot a_i}|},
\end{equation}
where $\eta_1$ is a timestep coefficient, $\mathbf{a_i}$ and $\mathbf{\dot a_i}$ are the acceleration of particle $i$ and its time-derivative, respectively.
In the case of the soft-sphere model, the timestep has to be much smaller than the impact duration:
\begin{equation}
\Delta t=\min(\eta_1 \min_i \frac{|\mathbf{a_i}|}{|\mathbf{\dot a_i}|}, \eta_2 T_s)
\end{equation}
where $\eta_2$ is a timestep coefficient.

\subsection{Disk Model \label{sec:model}}
We assume that all particles have the same mass $m_\mathrm{p}$.
The system parameters are the distance from the Sun $a_0$, the length of region $L$, the number of particles $N$, and the mass of particles $m_\mathrm{p}(r_\mathrm{p},\rho_\mathrm{p})$, where $\rho_\mathrm{p}$ is the material density of dust particles.
The dynamical behavior is characterized by only two non-dimensional parameters \citep{Daisaka1999}, the dynamical optical depth $\tau$ and the ratio $\zeta=r_\mathrm{H}/2r_\mathrm{p}$.
In the case where $\zeta>1$, the center of the particle is in the Hill sphere of another particle when two particles come into contact with each other.
The dynamical optical depth is given by
\begin{equation}
\tau=\frac{3\Sigma_\mathrm{d}}{4 \rho_\mathrm{p} r_\mathrm{p}}=0.19 \left(\frac{\Sigma_\mathrm{d}}{10 \mathrm{g\,cm}^{-2}} \right)
\left(\frac{\rho_\mathrm{p}}{2 \mathrm{g\,cm}^{-3}} \right)^{-1}
\left(\frac{r_\mathrm{p}}{20 \mathrm{cm}} \right)^{-1}.
\end{equation}
The ratio $\zeta$ is given by
\begin{equation}
\zeta=105.528 \left( \frac{M_\odot}{M_* }\right)^{-1/3} \left( \frac{\rho_\mathrm{p}}{2 \mathrm{g}\mathrm{cm}^{-3}}\right)^{1/3}
\left( \frac{a_0}{1 \mathrm{AU}}\right).
\end{equation}

Using the above two parameters, the other parameters can be estimated.
The normalized critical wavelength is 
\begin{equation}
\tilde \lambda_{\mathrm{most}}=12 \pi \tau \zeta^2.
\end{equation}
The normalized size of computational domain $\tilde L$, the number of particle $N$, the normalized radius of particle $\tilde r_\mathrm{p}$, and Toomre's $Q$ are given by
\begin{equation}
\tilde L = 12 \pi A \tau \zeta^2,
\end{equation}
\begin{equation}
N=576 \pi A^2 \tau^3 \zeta^6,
\label{eq:N}
\end{equation}
\begin{equation}
\tilde r_\mathrm{p} = \frac{1}{2 \zeta},
\end{equation}
\begin{equation}
Q= \frac{\tilde \sigma_x}{6 \tau \zeta^2},
\label{eq:qq}
\end{equation}
where $A$ is the ratio $L/\lambda_{\mathrm{most}}$ and $\tilde \sigma_x$ is the initial radial velocity dispersion scaled by $r_\mathrm{H} \Omega_0$.
In order to determine the size of the computational domain, we have to set $A$. The initial velocity dispersion is also a parameter. 
We determine the initial velocity dispersion from the initial Toomre's $Q_\mathrm{init}$ value using Equation (\ref{eq:qq}).
As shown above, if we set $\tau$, $\zeta$, $A$, and $Q_\mathrm{init}$, we can determine the initial state of the system.

From Equation (\ref{eq:N}), if we set $\zeta=105.528$ at 1AU, the number of particles becomes too large to handle with the currently available computers.
We have to use smaller $\zeta$.
According to \citet{Ohtsuki1993} and \citet{Salo1995}, the possibility of accretion is determined by the ratio $\zeta$.
Because the capture probability decreases abruptly for $\zeta < 3/2$, the ratio $\zeta$ should be larger than $3/2$.
Therefore we decided to set $\zeta=2$ in our simulation. 
The effect of $\zeta$ is discussed in \S \ref{sec:hr}.

\subsection{Initial Condition}
In the Hill coordinates, the Keplerian motion has six parameters: the position of the guiding center $(x_\mathrm{G},y_\mathrm{G})$, inclination $i$, eccentricity $e$, and two phases of motion \citep{Nakazawa1988}.
The eccentricity $e$ and inclination $i$ of particles are assumed to obey the Reyleigh distribution with the ratio $\langle e^2 \rangle ^{1/2}/\langle i^2 \rangle ^{1/2}=2$ \citep{Ida1992}. 
The eccentricity $e$ is related to the radial velocity dispersion $\langle e^2 \rangle ^{1/2}=\sqrt{2}\tilde \sigma_x h $.
The velocity dispersion is determined by Equation (\ref{eq:qq}). 
The position of guiding center $(x_\mathrm{G},y_\mathrm{G})$ and the phases of epicyclic and vertical motions are chosen randomly, avoiding overlapping of particles.  

The initial conditions are summarized in Table $\ref{table:model}$.
Models 1H and 1 are the fiducial models.
In model 1H the parameters are $\tau=0.1$, $\zeta=2.0$, $Q_\mathrm{init}=3.0$, and $A=6.0$.
The collision model is the hard-sphere model.
Model 1 has the same parameters as model 1H but its collision model is the soft-sphere model, and
the duration of collision is $0.02 T_\mathrm{K}$.

We can calculate the actual length scales. 
Using minimum mass solar nebular model the most unstable wavelength is $3.3 \times 10^3\mathrm{km}$. 
In the case of the fiducial model, the size of the computational domain is $2.0 \times 10^4 \mathrm{km}$.  
If the gaseous disk is turbulent, the vertical structure of dust particles is determined by the balance between the sedimentation due to gravity and diffusion due to turbulence.
The thickness is estimated as $H_\mathrm{d}/H_\mathrm{g}=\alpha/(\Omega_\mathrm{K} \tau_\mathrm{f})$ where $\alpha$ is a non-dimensional factor of the turbulent viscosity, $\tau_\mathrm{f}$ is the stopping time due to the friction \citep{Dubrulle1995}. 
We assume that $\alpha$ is very small as $\alpha\simeq 10^{-3}$. Assuming $D=20\mathrm{cm}$, the equilibrium thickness of a dust layer is $5.6 \times 10^5 \mathrm{km}$. 
We take the initial thickness of the dust layer in our simulation is about $\sigma_x/\Omega_\mathrm{K}=1.5\times10^3 \mathrm{m}$ in the case of the fiducial model. The thickness is thinner than the realistic value in order to reduce the calculation time.

The particle used in this simulation is a ``super particle" that represents the group of small dust particles at 1AU, i.e., the size of particles in this simulation is much larger than that of the actual particles and their internal density is much lower than that of actual particles. 
We obtain the radius $r_\mathrm{p}=5.9\times 10^6 \mathrm{cm}$ and the density $\rho_{\mathrm{p}}=1.4\times 10^{-5} \mathrm{g}\mathrm{cm}^{-3}$ from $\zeta=2.0$, and $\tau=0.1$.
One super particle represents $1.5 \times 10^{11}$ dust particles.
We may think that some dust particles in the super particles do not collide when the super particles collide.
In addition, we may also think that the angles and the relative velocities of the collisions of the individual particles in a collision of two super particles may vary much.
Thus the restitution coefficient $\epsilon$ used in this simulation does not represent the physical restitution coefficient, but corresponds to the coefficient of the dissipation due to collisions between super particles. 
\begin{deluxetable}{cccccccc}
\tablecaption{Initial conditions for dust disks}
\tablewidth{0pt}
\tablehead{
\colhead{Model} & \colhead{$\epsilon$} & \colhead{$\tau$} & \colhead{$\zeta$} & \colhead{	$A$} &
\colhead{$Q_\mathrm{init}$} & \colhead{Collision} & \colhead{$T_s$} 
}
\startdata 
1H   & 0.01       & 0.1    & 2.0     & 6.0 & 2.92 & Hard             &      \\
1    & 0.01       & 0.1    & 2.0     & 6.0 & 2.92 & Soft            & 0.02   \\
2   & 0.01       & 0.1    & 2.0     & 6.0 & 2.92 & Soft            & 0.09   \\
3   & 0.01       & 0.1    & 2.0     & 6.0 & 2.92 & Soft            & 0.05  \\
4  & 0.01       & 0.1    & 2.0     & 6.0 & 0.42 & Soft           & 0.02 \\
5  & 0.01       & 0.05   & 2.0     & 6.0 & 5.83 & Soft           & 0.02 \\
6  & 0.01       & 0.125  & 2.0     & 6.0 & 2.33 & Soft           & 0.02 \\
7  & 0.1        & 0.1    & 2.0     & 6.0 & 2.92 & Soft           & 0.02 \\
8  & 0.3        & 0.1    & 2.0     & 6.0 & 2.92 & Soft           & 0.02 \\
9  & 0.5        & 0.1    & 2.0     & 6.0 & 2.92 & Soft           & 0.02 \\
10  & 0.7        & 0.1    & 2.0     & 6.0 & 2.92 & Soft           & 0.02 \\
11  & 0.9        & 0.1    & 2.0     & 6.0 & 2.92 & Soft           & 0.02 \\
12  & 0.01       & 0.05   & 2.0     & 3.0 & 3.33 & Soft           & 0.02 \\
13  & 0.01       & 0.05   & 2.0     & 6.0 & 3.33 & Soft           & 0.02 \\
14  & 0.01       & 0.05   & 2.0     & 8.0 & 3.33 & Soft           & 0.02 \\
15  & 0.01       & 0.05   & 2.0     & 10.0 & 3.33 & Soft           & 0.02 \\
16  & 0.01       & 0.05   & 2.0     & 12.0 & 3.33 & Soft           & 0.02 \\
17  & 0.01       & 0.05   & 1.75    & 6.0 & 7.62 & Soft           & 0.02 \\
18  & 0.01       & 0.05   & 2.0     & 6.0 & 5.83 & Soft           & 0.02 \\
19  & 0.01       & 0.05   & 3.0     & 6.0 & 2.60 & Soft           & 0.02 \\
\enddata

\tablecomments{$\epsilon$, $\tau$, $\zeta$, $A$, $T_s$, and $Q_\mathrm{init}$ are the restitution coefficient, the optical depth, the ratio $r_\mathrm{H}/2r_\mathrm{p}$, $L/\lambda_\mathrm{most}$, the duration of collision, and the initial Toomre's $Q$, respectively. 
}
\label{table:model}
\end{deluxetable}

\subsection{Test Calculation}
\begin{figure}
\plotone{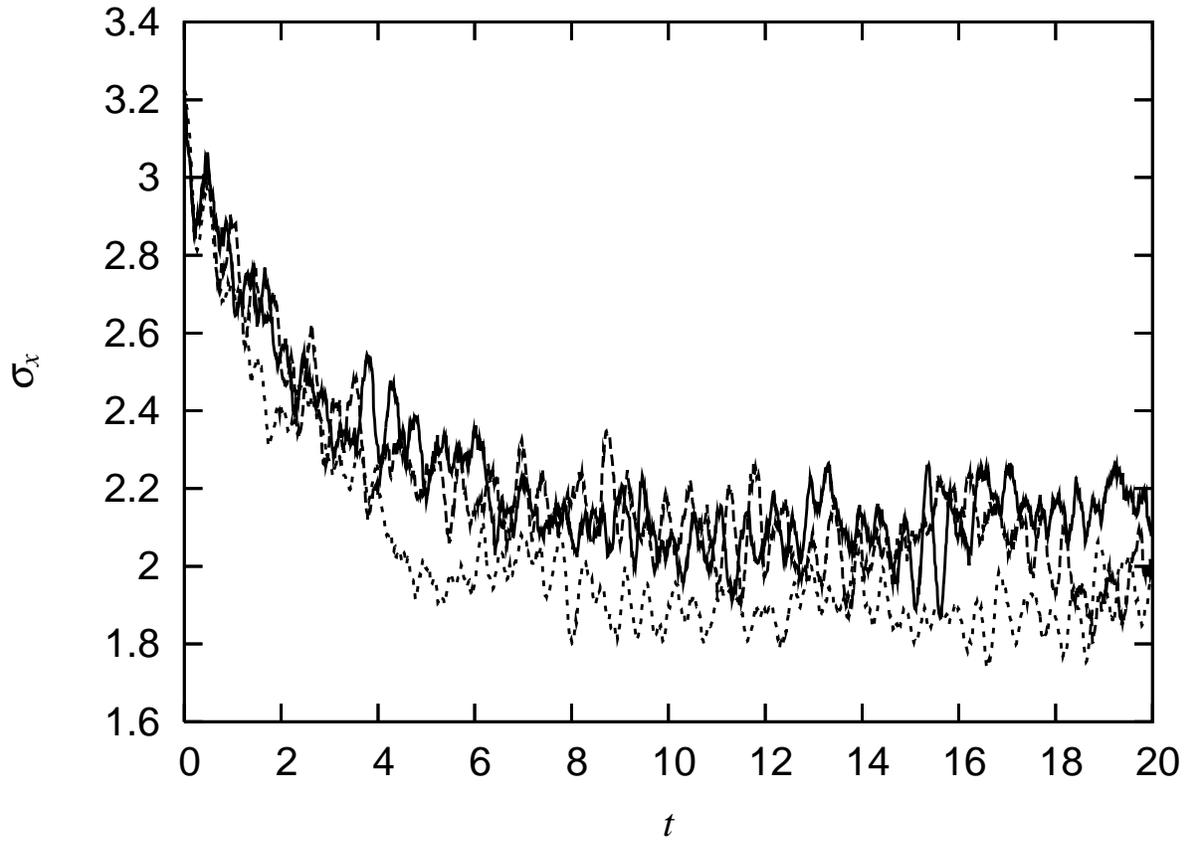}
\caption{The example of the time evolution of the velocity dispersion of the hard-sphere model (\textit{solid curve}), 
the soft-sphere models with $T_s=0.00053$ (\textit{dashed curve}), and $T_s=0.027$ (\textit{dotted curve}). The parameters are $\epsilon=0.5$, $\tau=0.1$, and $A=27$.
}
\label{fig:relax_soft.eps} 
\end{figure}

We perform the simulations with the same parameters as \citet{Salo1995} and \citet{Daisaka1999} in order to check the accuracy of our code. We calculate the radial velocity dispersion.
Each simulation lasts $20T_\mathrm{K}$ and we calculate the average and the standard deviation during the last $5 T_\mathrm{K}$.
Figure \ref{fig:relax_soft.eps} shows the time evolution of the velocity dispersion for the hard-sphere and soft-sphere models.
The velocity dispersion seems to approach a certain limit value. 
For the parameter $\zeta=0.82$, $\tau=0.1, 0.4, 0.6$, and $\epsilon=0.1, 0.4, 0.6, 0.7$, the equilibrium states exist. 
Table \ref{table:vel} shows that the velocity dispersions of our calculation with the soft-sphere and hard-sphere models agree with those of \citet{Salo1995} and \citet{Daisaka1999}. 

The equilibrium value of the soft-sphere model must approach that of the hard-sphere model within the limit of the large stiffness of the spring $k \to \infty$.
We set the parameters as $\zeta=0.82$, $\tau=0.1$, and $\epsilon=0.5$.
Figure \ref{fig:relax_soft.eps} shows that there is little difference between the results of the hard-sphere and soft-sphere models if the duration $T_s$ is sufficiently short.
The difference is prominent if the aggregates are formed
(See \S \ref{sec:softhard}).
In the case of these parameters, aggregates are not formed, thus we find no remarkable difference.

\begin{deluxetable}{ccccccc}
\tablecaption{Equilibrium radial velocity dispersion}
\tablewidth{0pt}
\tablehead{
\colhead{$\epsilon$} & \colhead{$\tau$} & \colhead{$ A$} & 
\colhead{Hard} & 
\colhead{Soft} &
\colhead{Salo (1995)} & \colhead{Daisaka \& Ida (1999)}
}
\startdata
$0.50$ & $0.40$ & $6.80$ & $2.99 \pm 0.27$ & $3.05 \pm 0.30$ & $3.01 \pm 0.27 $ & $2.91 \pm 0.26$ \\
$0.50$ & $0.60$ & $4.40$ & $5.06 \pm 1.13$ & $5.30 \pm 1.18$ & $5.84 \pm 1.42 $ & $5.09 \pm 1.14$ \\
$0.10$ & $0.40$ & $5.12$ & $4.10 \pm 0.83$ & $4.34 \pm 0.99$ & $4.39 \pm 0.67 $ & $4.50 \pm 0.59$ \\
$0.40$ & $0.40$ & $5.12$ & $3.19 \pm 0.55$ & $3.30 \pm 0.54$ & $3.48 \pm 0.43 $ & $4.06 \pm 0.59$ \\
$0.60$ & $0.40$ & $5.12$ & $2.53 \pm 0.15$ & $2.47 \pm 0.18$ & $2.61 \pm 0.18 $ & $2.58 \pm 0.17$ \\
$0.70$ & $0.40$ & $5.12$ & $3.94 \pm 0.10$ & $3.62 \pm 0.17$ & $3.60 \pm 0.07 $ & $3.67 \pm 0.11$ \\  
\enddata
\label{table:vel}
\end{deluxetable}

\clearpage

\section{Results \label{sec:res}}

\subsection{Formation Process of Planetesimals}
We perform the simulation with the various models (see Table \ref{table:model}).
We find that the formation process of planetesimals is qualitatively the same in all models, except for the model where the initial velocity dispersion is small ($Q_\mathrm{init}<1$), namely, where a dust layer is initially unstable.

\begin{figure}
  \begin{center}
    \begin{tabular}{ccc}
      \resizebox{50mm}{!}{\includegraphics{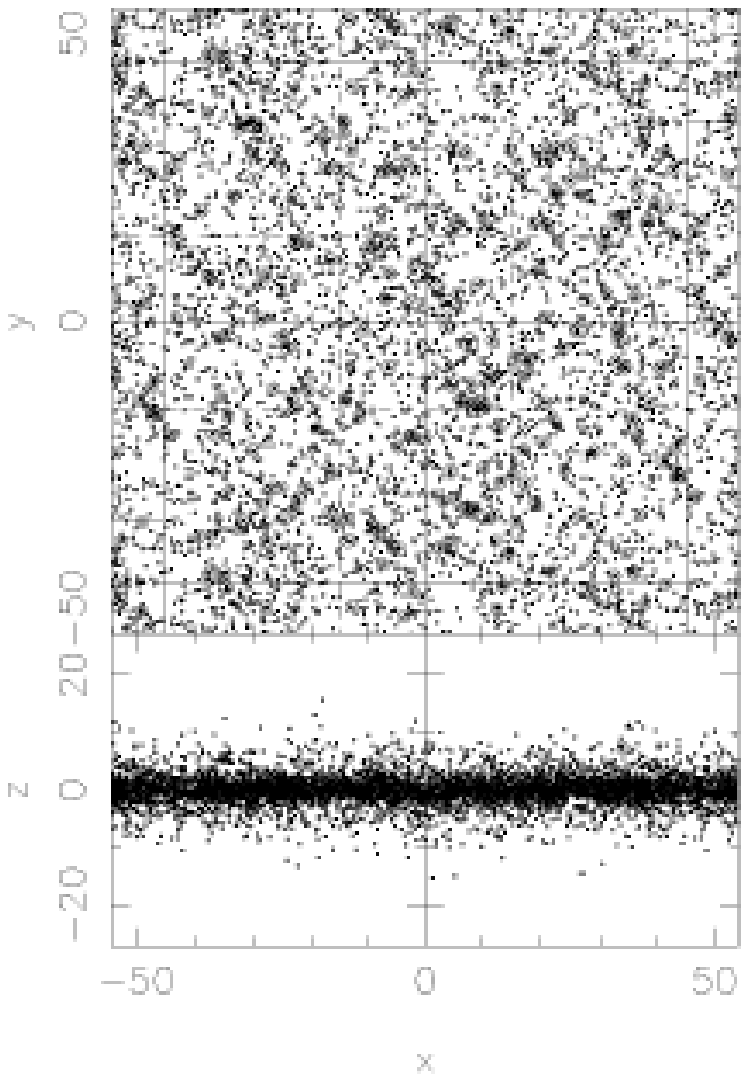}} &
      \resizebox{50mm}{!}{\includegraphics{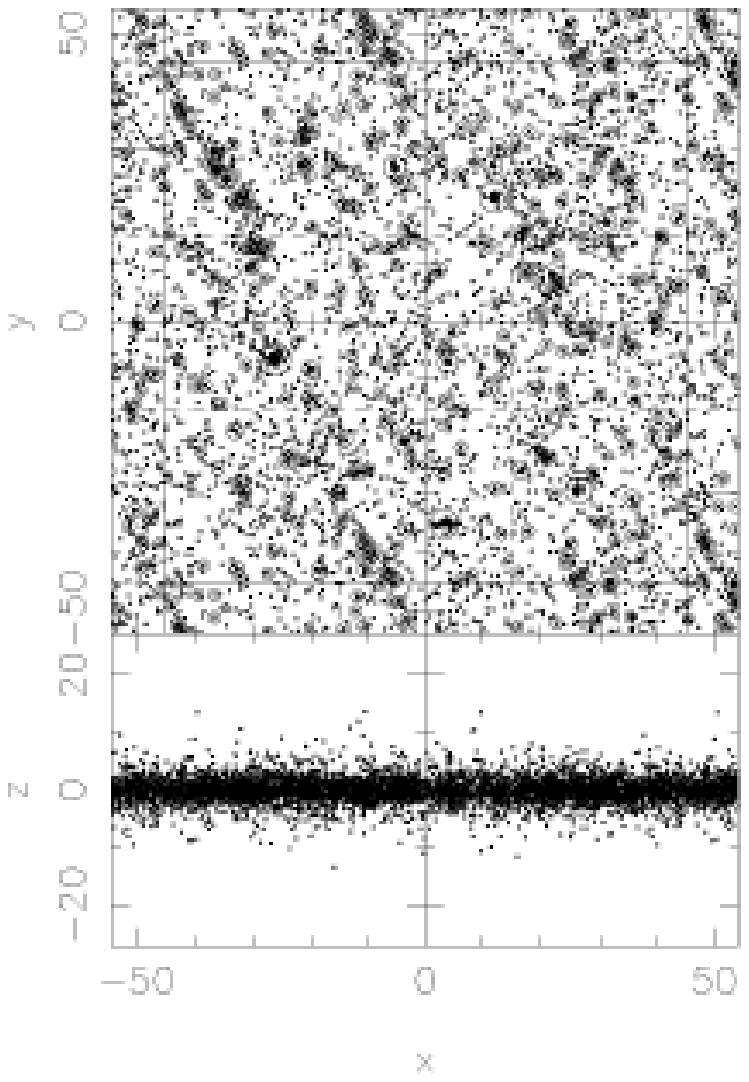}} &
      \resizebox{50mm}{!}{\includegraphics{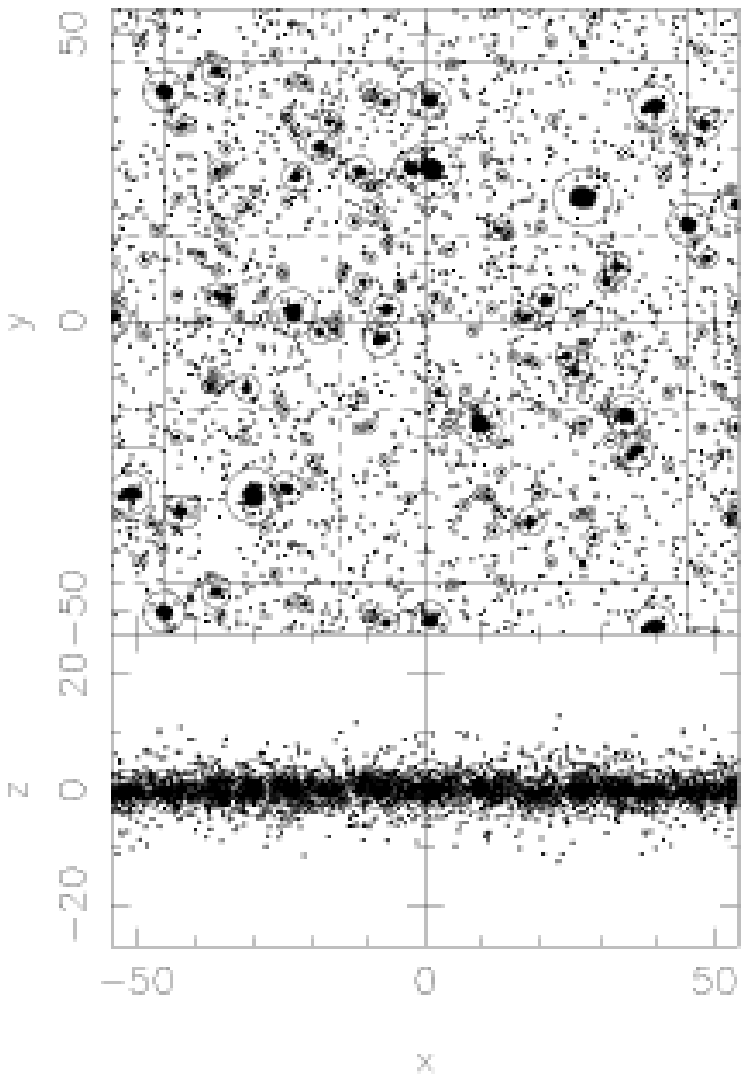}} \\
      \resizebox{50mm}{!}{\includegraphics{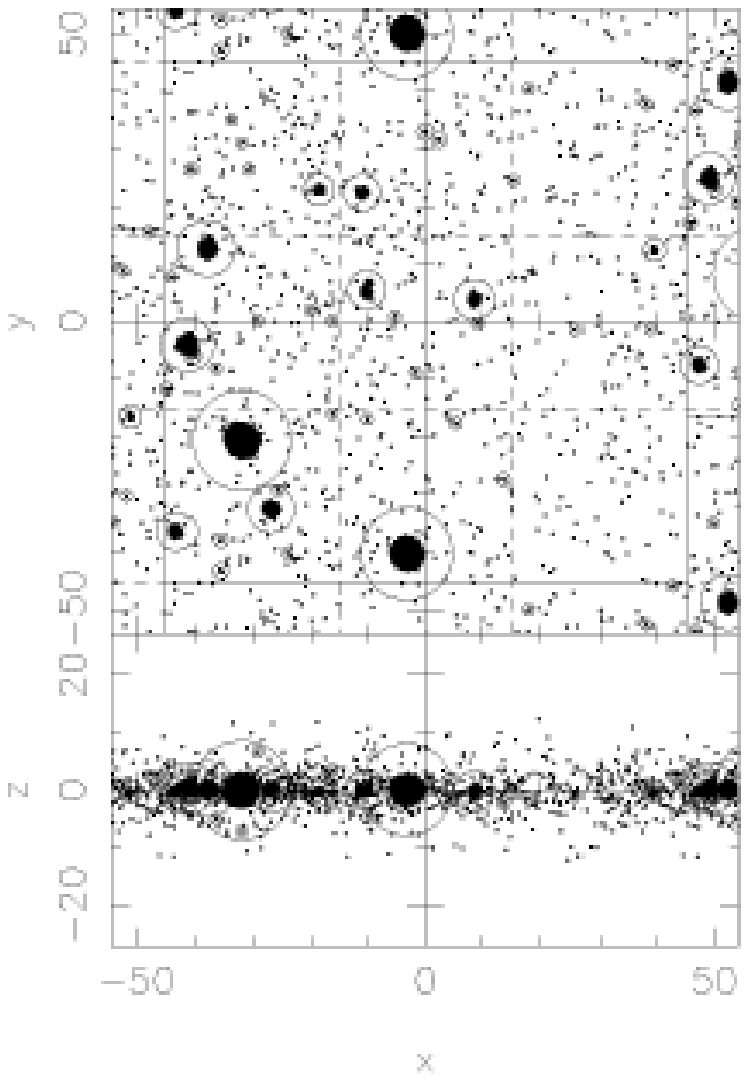}} &
      \resizebox{50mm}{!}{\includegraphics{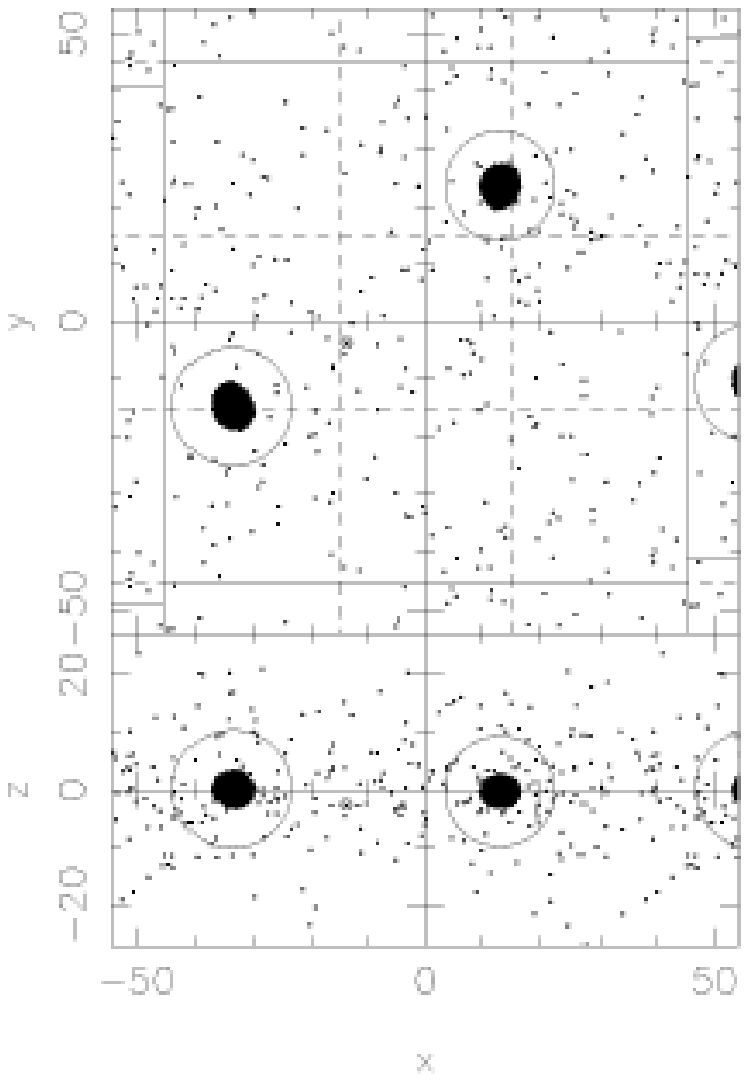}} &
      \resizebox{50mm}{!}{\includegraphics{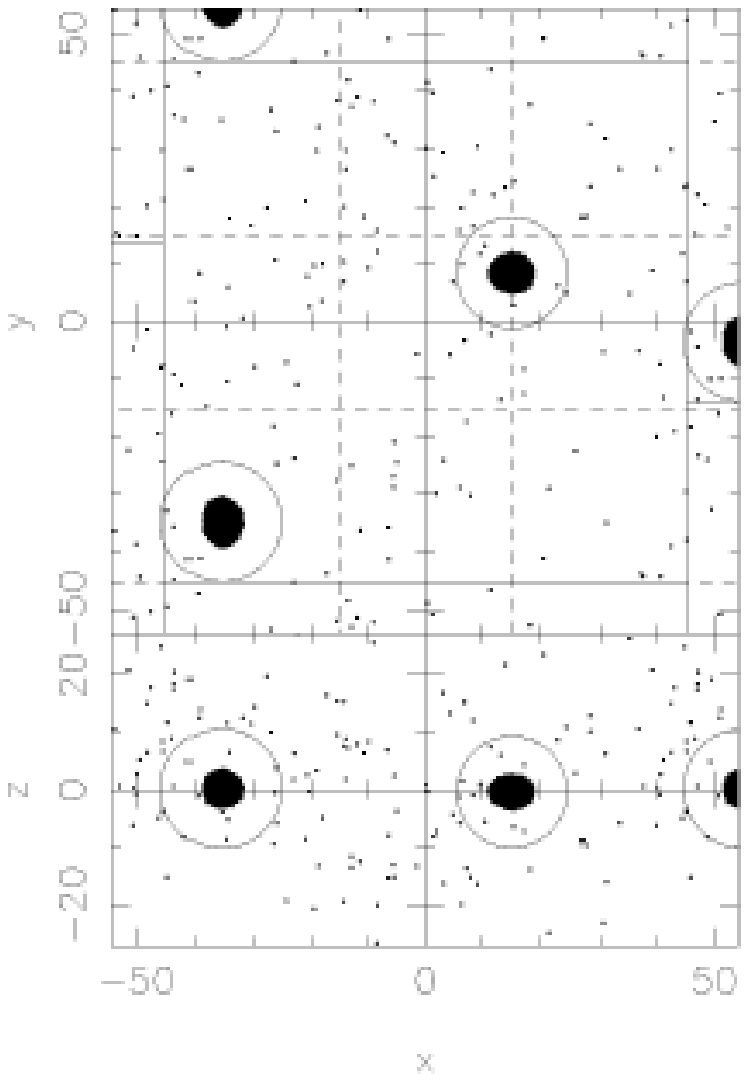}} \\      
    \end{tabular}
\caption{The spatial particle distribution in Model 1 at $t=0.5 T_\mathrm{K}$ (\textit{top left panel}), $t=1.0 T_\mathrm{K}$ (\textit{top middle panel}), $t=2.0 T_\mathrm{K}$ (\textit{top right panel}),
$t=3.0 T_\mathrm{K}$ (\textit{bottom left panel}), $t=6.0 T_\mathrm{K}$ (\textit{bottom middle panel}), and $t=10.0 T_\mathrm{K}$ (\textit{bottom right panel}). Solid circles denote the Hill sphere of aggregates.
}
\label{fig:run07pic}
  \end{center}
\end{figure}

First, we explain the overall formation process of planetesimals through GI in the case where $Q_\mathrm{init}>1$.
Figure \ref{fig:run07pic} shows snapshots of model 1 at $t/T_\mathrm{K}=0.5 , 1.0 , 2.0 , 3.0 , 6.0 ,$ and $10.0$.
In this run, the velocity dispersion of particles is initially large enough for the layer to be stable ($Q_\mathrm{init}=2.9$). 
The velocity dispersion of particles decreases due to inelastic collisions, the dust layer becomes thinner and lessens Toomre's $Q$ value.
The non-axisymmetric wake-like structure develops gradually when the velocity dispersion becomes sufficiently small ($Q\simeq 2$).
The wake-like structure has been examined by many authors \citep[e.g.][]{Salo1995,Daisaka1999}.
The wakes are created if we include both self-gravity and inelastic collisions.
The dense parts of wakes fragment into aggregates due to GI. 
The time scale of the fragmentation is the Kepler period.
The mass of aggregates is discussed in \S \ref{sec:mass}.
These aggregates correspond to planetesimals.
Once the aggregates are formed, they grow rapidly through fast mutual coalescences.
In summary, the process of planetesimal formation can be divided into three stages: the formation of the non-axisymmetric wake-like structure, the formation of aggregates, and the rapid collisional growth of aggregates.
The formation process for the hard-sphere model (model 1H) is the same as that for the soft-sphere model.

\begin{figure}
  \begin{center}
    \begin{tabular}{ccc}
      \resizebox{50mm}{!}{\includegraphics{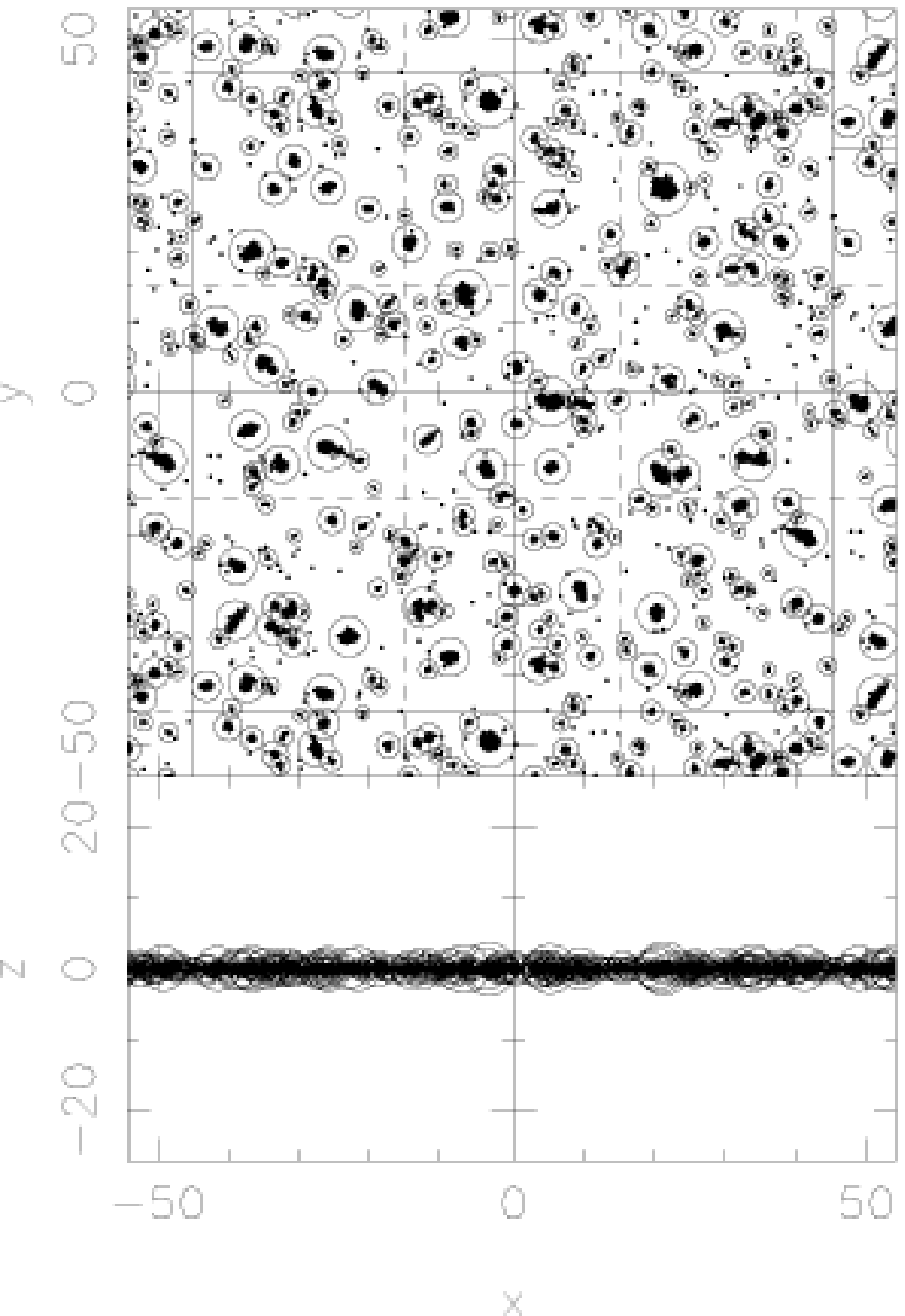}} &
      \resizebox{50mm}{!}{\includegraphics{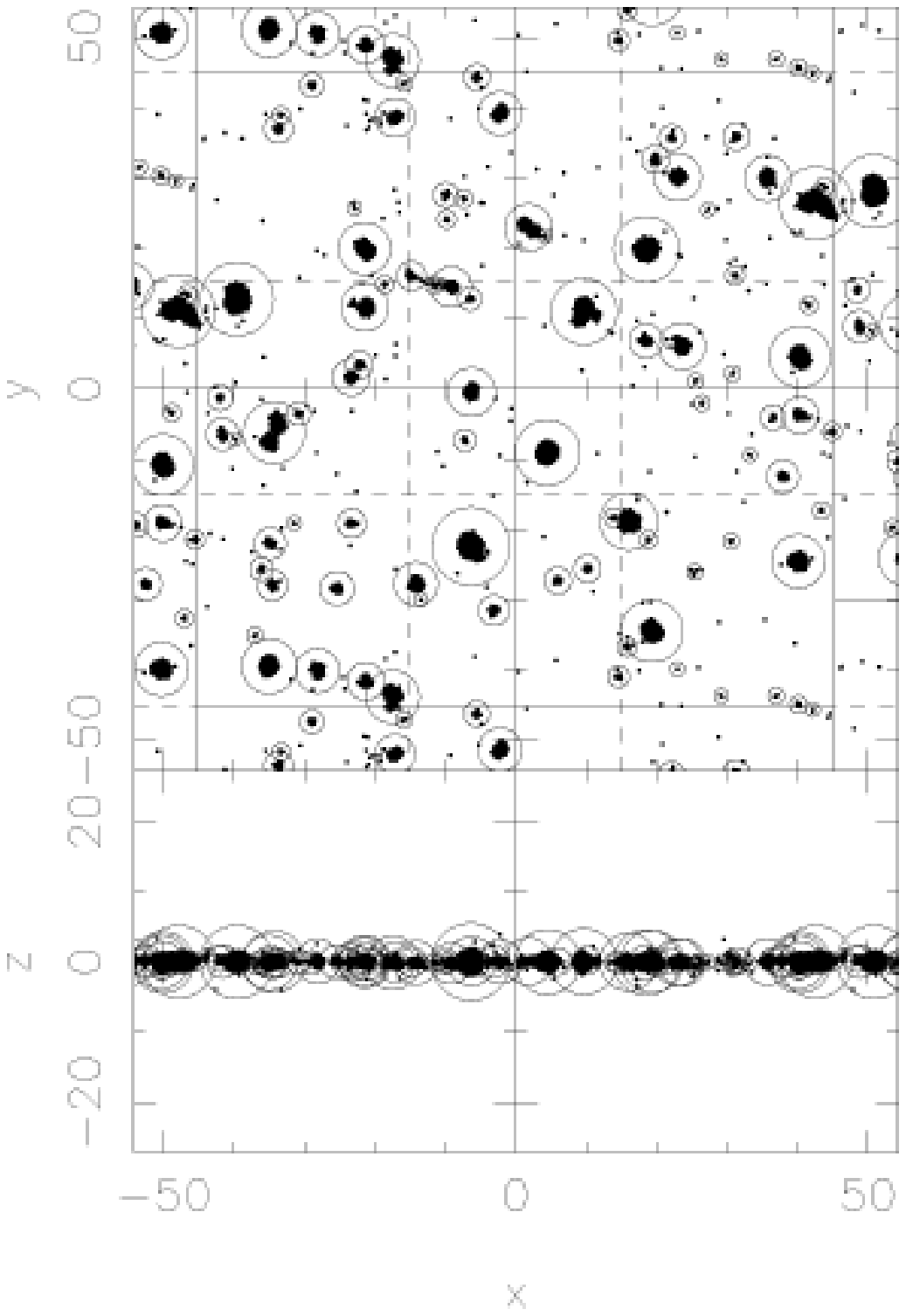}} &
      \resizebox{50mm}{!}{\includegraphics{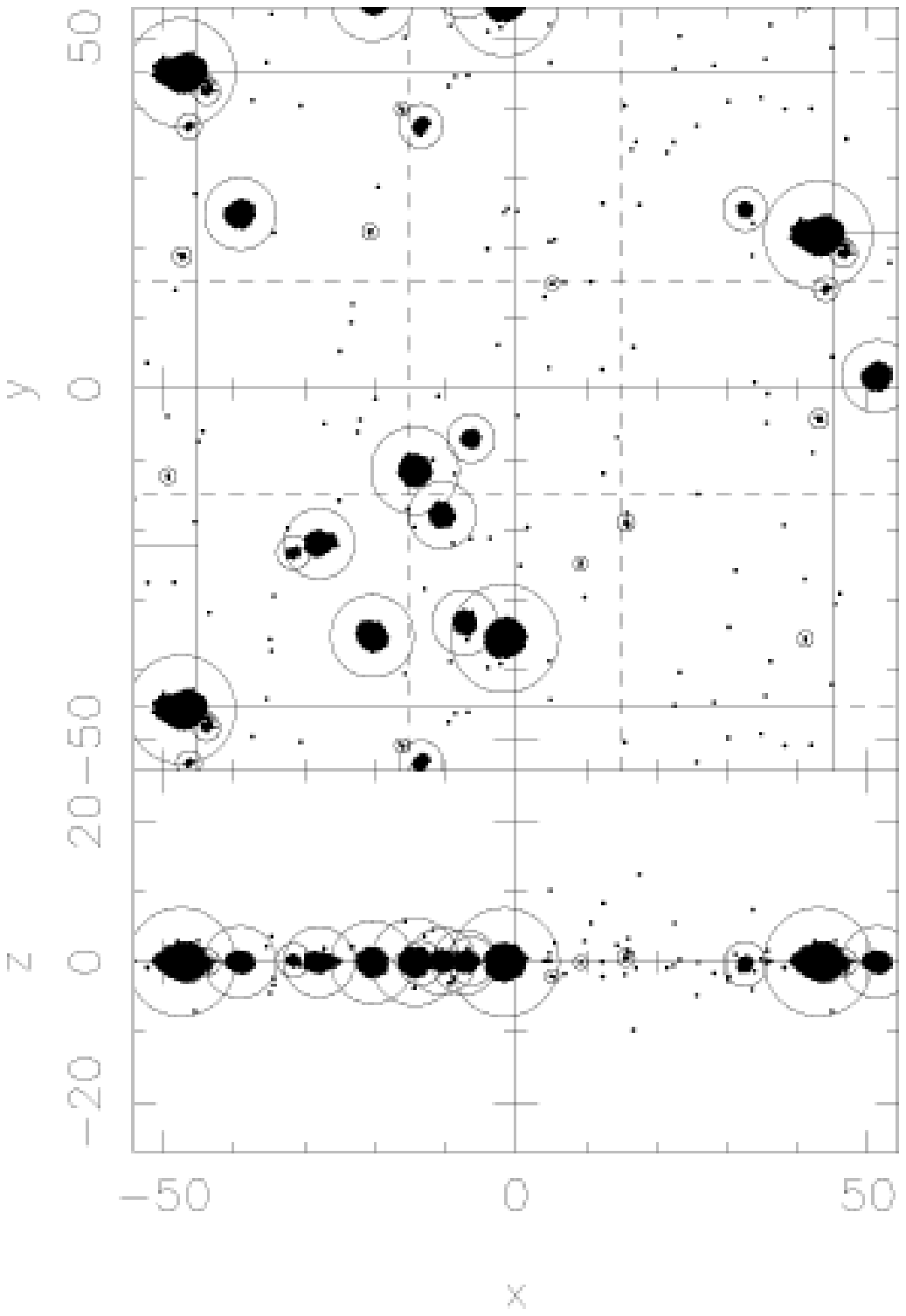}} \\
      \resizebox{50mm}{!}{\includegraphics{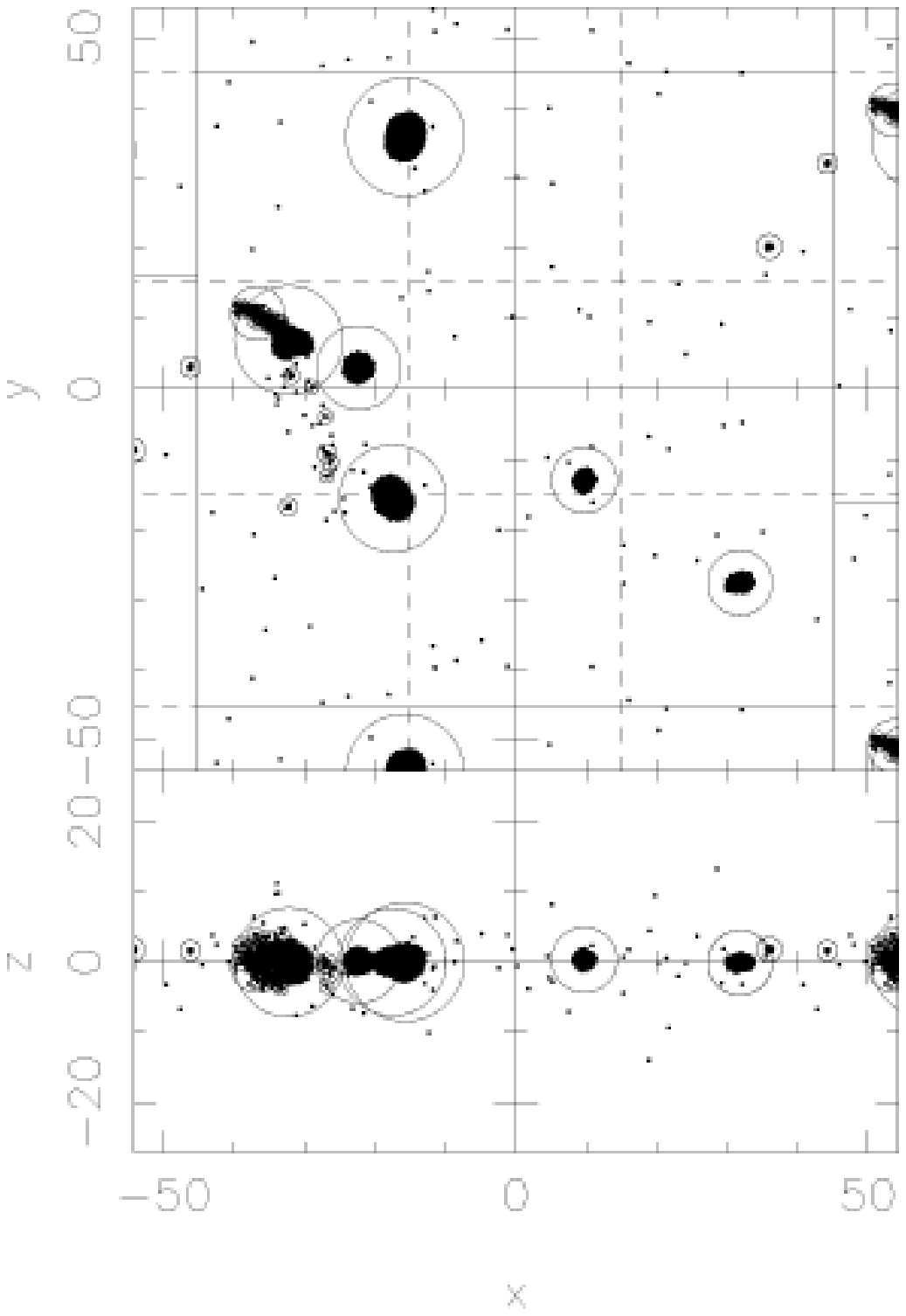}} &
      \resizebox{50mm}{!}{\includegraphics{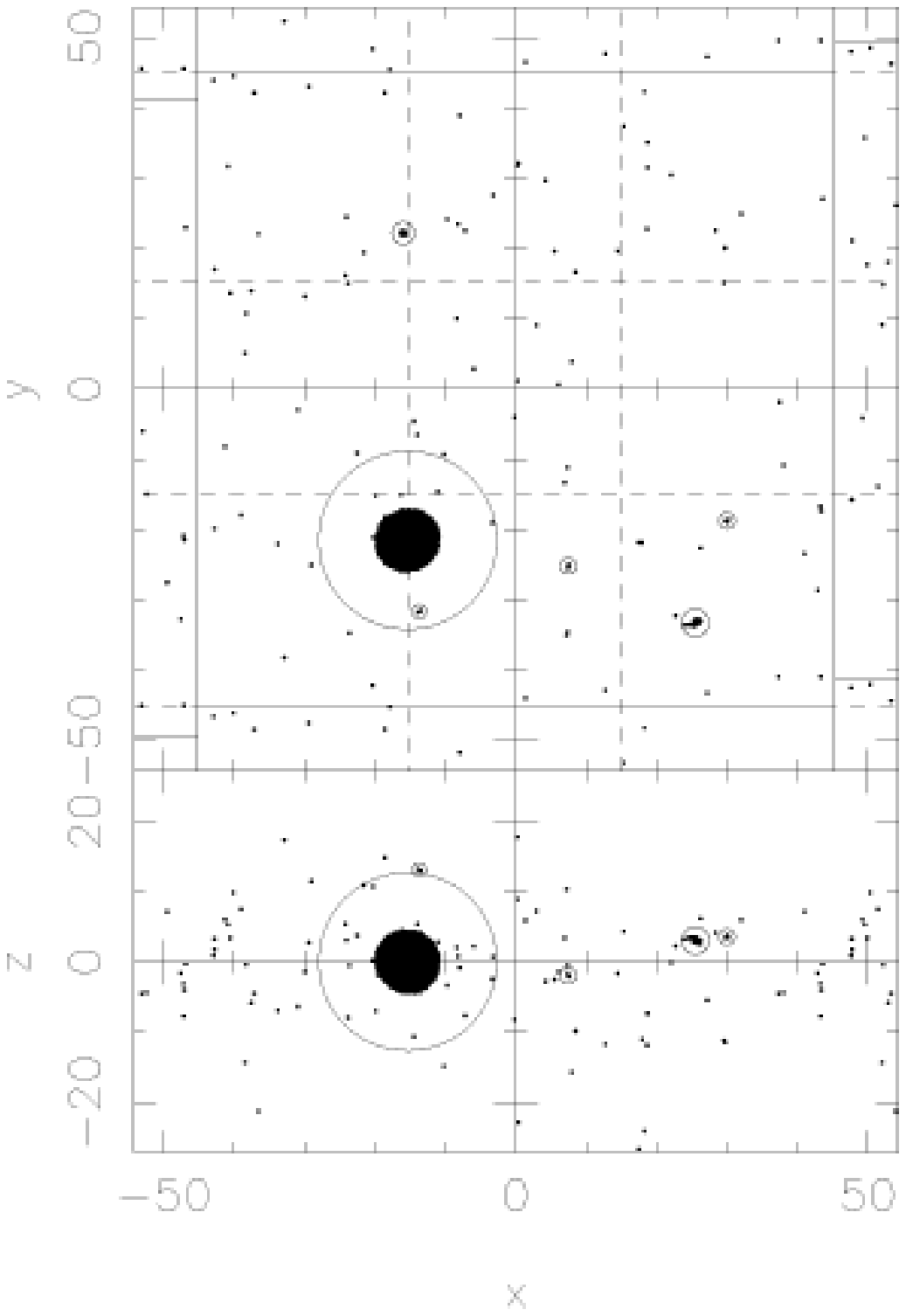}} &
      \resizebox{50mm}{!}{\includegraphics{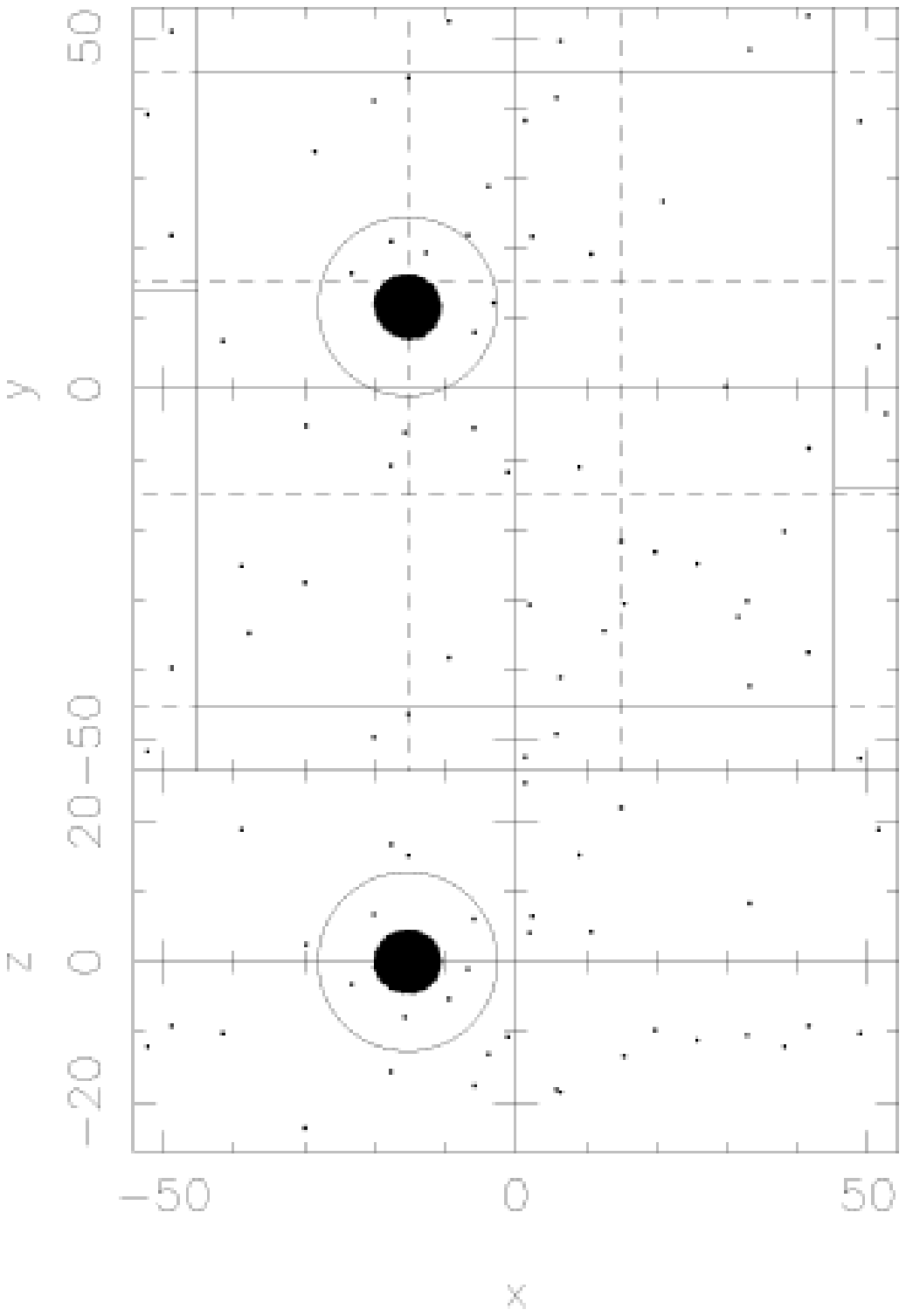}} \\      
    \end{tabular}
\caption{
Same as the Figure \ref{fig:run07pic} but for the case of $Q_\mathrm{init}=0.42$ (model 4).
}
\label{fig:run110pic}
  \end{center}
\end{figure}

The planetesimal formation in the case with $Q_\mathrm{init}<1$ is different from the case with $Q_\mathrm{init}>1$ as shown in Figure \ref{fig:run110pic}.
The particle distribution imediately becomes non-uniform, and the non-axisymmetric wake structure is not formed. 
The denser regions of the non-uniform structure collapse and directly grow up into aggregates.
The subsequent evolution is the same as that for $Q_\mathrm{init}>1$. 

The flow of the gas component may be turbulent in the initial stage of planetesimal formation because of Kelvin-Helmholtz instability or magnetorotational instability \citep{Balbus1991}. Dust grains are stirred by turbulent gas, thus their velocity dispersion might initially be large. Therefore, we think the case where $Q_\mathrm{init}>2$ is more realistic. In the following section, we examine the case where $Q_\mathrm{init}>2$.

\afterpage{\clearpage}  

\afterpage{\clearpage}  

\subsection{Time Evolution}
We investigate the time evolution in our fiducial model (models 1H and 1).
We describe the processes common to all models.
We show the time evolutions of mass distribution, number of aggregates, mass and radius of the largest aggregate, and the velocity dispersion.
We make a comparison between the hard-sphere and soft-sphere models.

\subsubsection{Mass Distribution of Planetesimals \label{sec:mass}}
We define aggregates by the method used by \citet{Furuya2004}. At first, we detect an aggregate of particles whose mutual distances are smaller than the Hill radius of each particle. These particles are strongly attracted by mutual gravitational force. However, some particles that have sufficiently large velocity cannot stay in the aggregate. We must impose the binding condition and remove gravitationally unbound particles. 
The binding condition is determined by the Jacobi energy \citep[e.g.][]{Nakazawa1988}:
\begin{equation}
J_i= \frac{1}{2}v^2 - \frac{3}{2} x^2 + \frac{1}{2}z^2 - \frac{3}{2} \sum_{j=1}^{n_{ic}} \frac{1}{r_{ij}}+\frac{9}{2} r_{\mathrm{Hc}}^2<0,
\end{equation}
where $v$ is the relative velocity between the aggregate's center of mass and $i$-th particle, $x$ and $z$ are the relative position between the aggregate's center of mass and $i$-th particle, $n_{ic}$ is the number of particles in the aggregate, and $r_{Hc}$ is the Hill radius of the aggregate.

\begin{figure}
  \begin{center}
    \begin{tabular}{ccc}
      \resizebox{50mm}{!}{\includegraphics{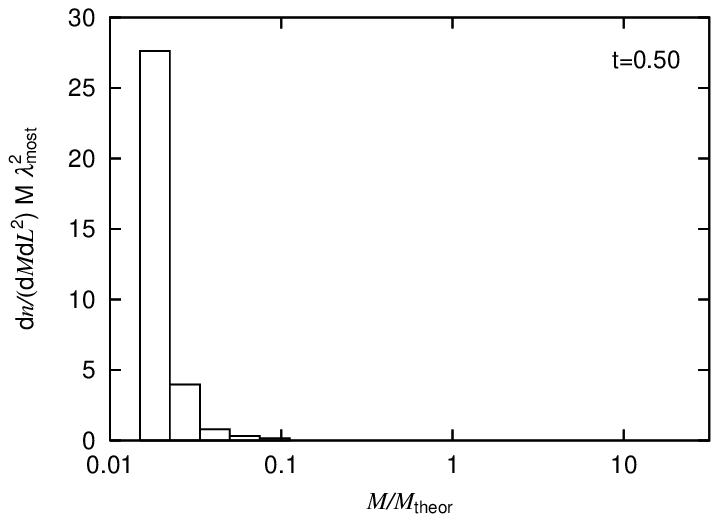}} &
      \resizebox{50mm}{!}{\includegraphics{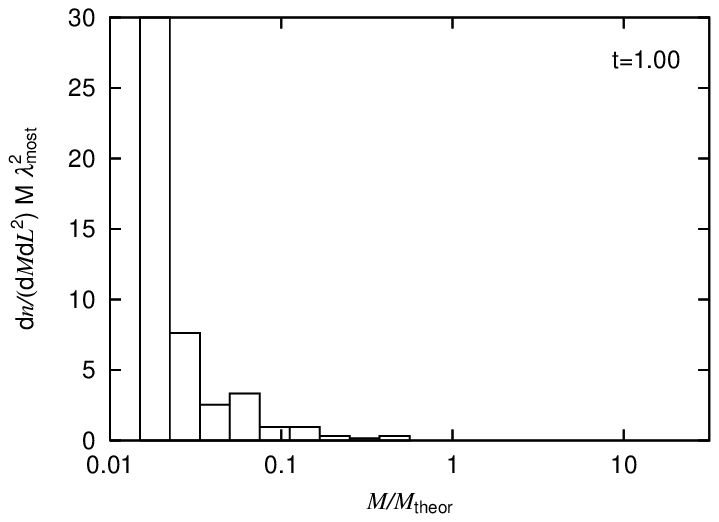}} &
      \resizebox{50mm}{!}{\includegraphics{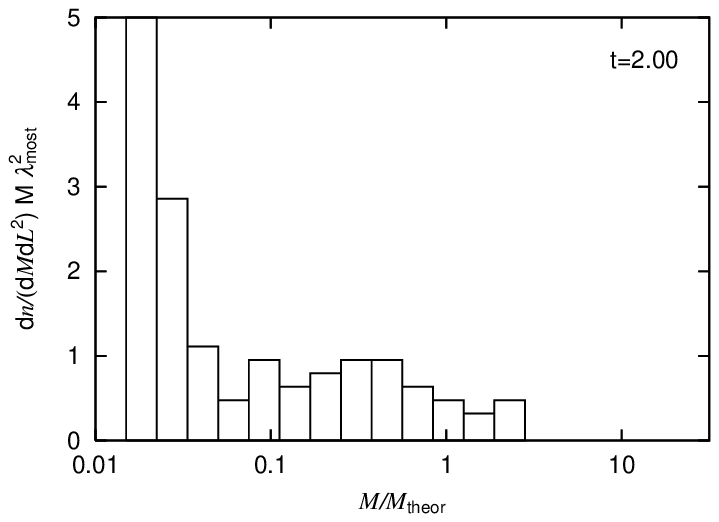}} \\
      \resizebox{50mm}{!}{\includegraphics{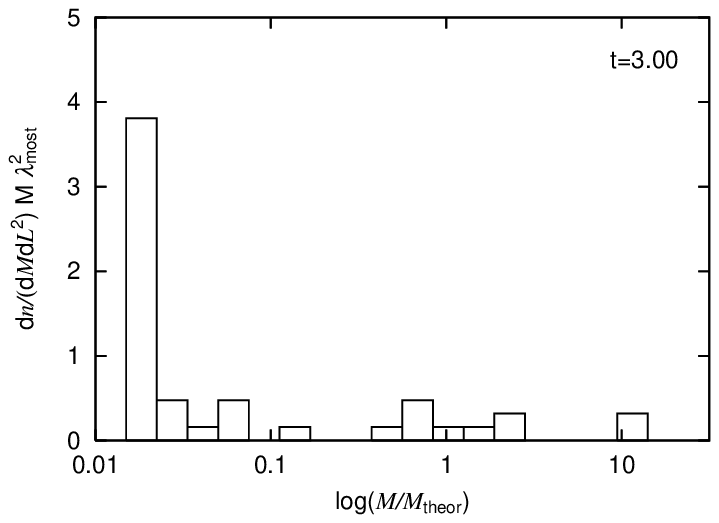}} &
      \resizebox{50mm}{!}{\includegraphics{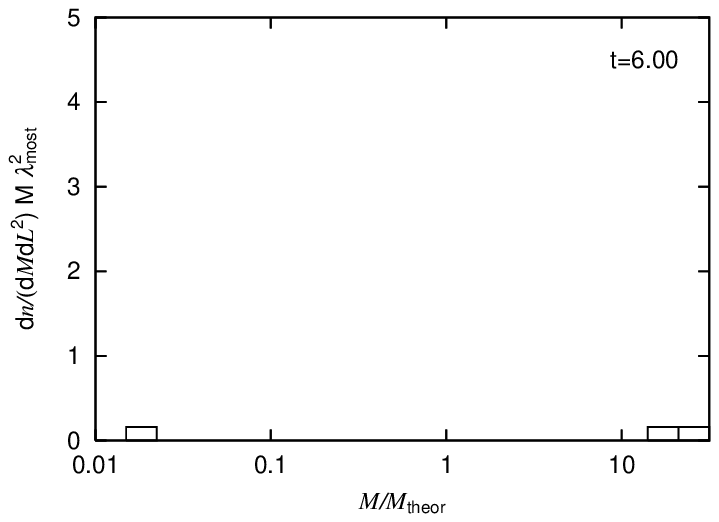}} &
      \resizebox{50mm}{!}{\includegraphics{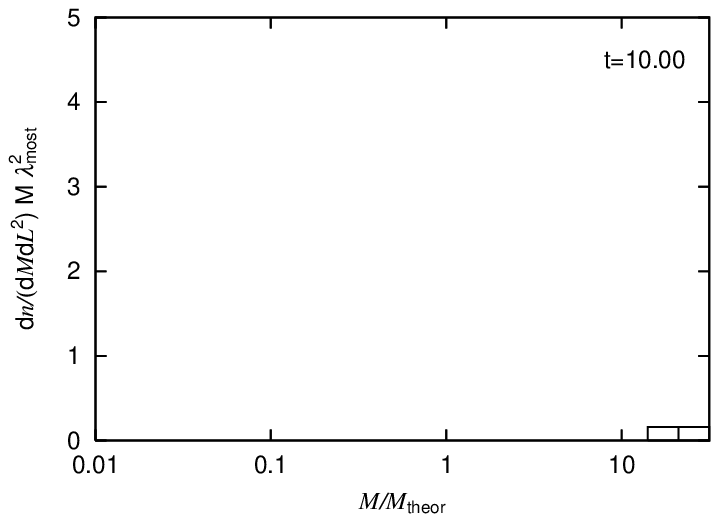}} \\      
    \end{tabular}
\caption{The snapshots of the mass distribution for model 1 at $t=0.5 T_\mathrm{K}$ (\textit{top left panel}), $t=1.0 T_\mathrm{K}$ (\textit{top middle panel}), $t=2.0 T_\mathrm{K}$ (\textit{top right panel}),
$t=3.0 T_\mathrm{K}$ (\textit{bottom left panel}), $t=6.0 T_\mathrm{K}$ (\textit{bottom middle panel}), and $t=10.0 T_\mathrm{K}$ (\textit{bottom right panel}). The vertical axis denotes the number of aggregates per unit mass and area.}
\label{fig:run07mass}
  \end{center}
\end{figure}

Figure \ref{fig:run07mass} shows the mass distribution for model 1.
The aggregate with mass $m>0.1m_{\mathrm{theor}}$ is not formed in $0\le t \le 0.3 T_\mathrm{K}$.  
However, we observe the non-axisymmetric wake-like structure at $t = 1.0 T_\mathrm{K}$. 
The wake-like structure is caused by GI. 
At $t=2 T_\mathrm{K}$ the aggregates with mass $0.1 - 1.0 M_{\mathrm{theor}}$ are formed.
The masses of these aggregates are on the same order of fragmentation mass estimated by the linear theory.
For $t>3 T_\mathrm{K}$ the number of aggregates decreases and the larger aggregates are formed by the coalescence of the aggregates. 
Finally, a few large aggregates and somewhat smaller aggregates are formed. The largest aggregate reaches about $12 M_{\mathrm{theor}}$ in this run at $t = 10.0 T_\mathrm{K}$.
The time evolution of the mass distribution for the hard-sphere model (model 1H) is the same as that for the soft-sphere model.

\afterpage{\clearpage}  

\subsubsection{Number of Aggregates}
\begin{figure}
  \begin{center}
    \begin{tabular}{cc}
      \resizebox{50mm}{!}{\includegraphics{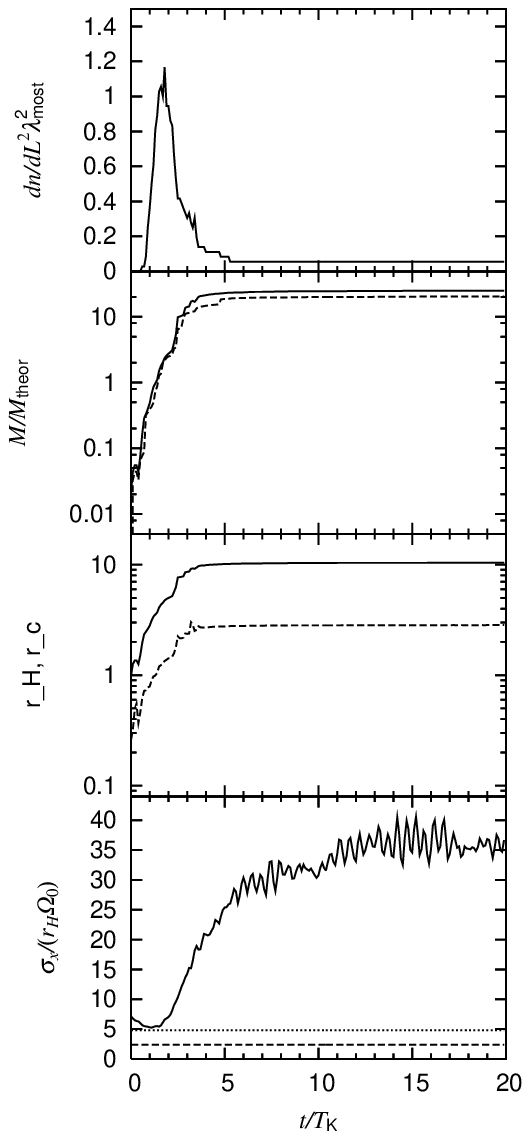}} &
      \resizebox{50mm}{!}{\includegraphics{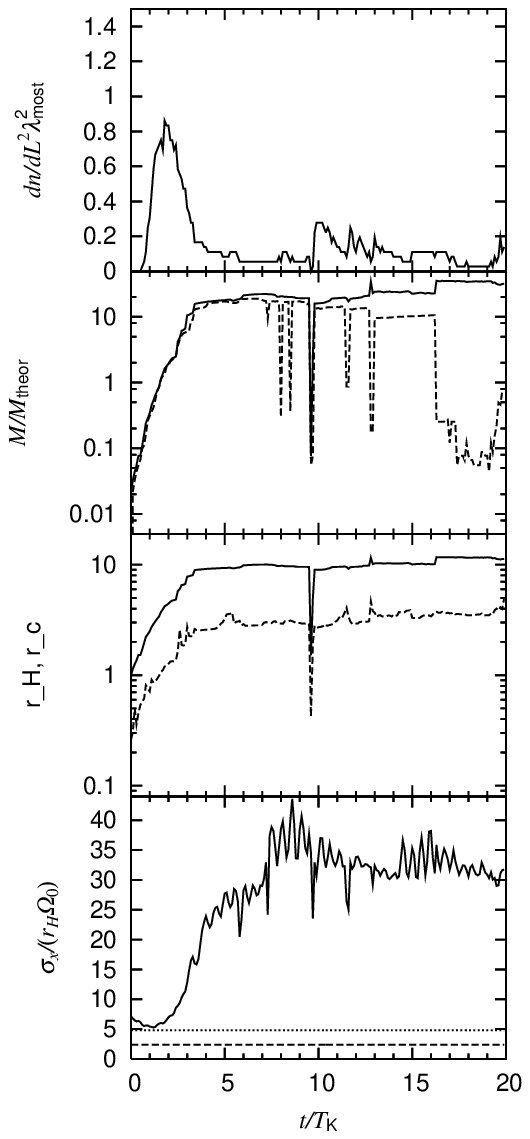}} \\
    \end{tabular}
\caption{The time evolution for the soft-sphere model (model 1) (\textit{left panel}) and the hard-sphere model (model 1H) (\textit{right panel}).
The quantities in this figure are the number of aggregates, the mass of the largest and second-largest aggregates, the Hill radius (\textit{solid curve}) and physical radius (\textit{dotted curve}) of the largest aggregate, and the velocity dispersion of field particles from top to bottom, respectively. In the bottom panel we show the velocities for $Q=2$ (\textit{dotted curve}) and $Q=1$ (\textit{dashed curve}).
}
\label{fig:all}
  \end{center}
\end{figure}

We cannot find any remarkable difference between the soft-sphere and hard-sphere models
in the time evolution of the number of aggregates in Figure \ref{fig:all}.
For $t<2 T_\mathrm{K}$, the number of aggregates increases monotonically up to about $1.2$ per $\lambda _\mathrm{most}^2$.
According to the linear theory, a few aggregates are formed per $\lambda _\mathrm{most}^2$. 
This result indicates that these aggregates correspond to planetesimals predicted in the linear theory.
However, the number of aggregates decreases rapidly after $t \simeq 2 T_\mathrm{K}$. 
This decay is caused by the fast coalescence of the aggregates. 
A few large aggregates absorb many other small aggregates, and thus the total number of aggregates decreases.

\subsubsection{The Largest Aggregate \label{sec:largestclump}}
The time evolution of the mass of the largest and second largest aggregates is shown in Figure \ref{fig:all}.
In the hard-sphere model (model 1H), the largest aggregates grow to about $8 m_{\mathrm{theor}}$ until $t=4T_\mathrm{K}$. 
There is little difference between the largest aggregate and the second-aggregate for $t<3T_\mathrm{K}$.
However, the mass of the second-largest aggregate decreases abruptly at $t\sim 3 T_\mathrm{K}$. 
This decrease is caused by the coalescence of the largest aggregate and the second-largest aggregates. 
A few aggregates absorbed almost mass in the computational domain.

At $t\simeq 9.6 T_\mathrm{K}$, there is a deep cusp, that is, the mass of the largest aggregate decreases abruptly, and restores immediately.
This is an artificial phenomenon due to the cluster finding method.
We see such a rapid decrease and restoration, when two large aggregates collide but do not finally merge into a single aggregate. 
When a collision between large aggregates with large relative velocity occurs, the particles of the transient object have high relative velocities, so the binding conditions for the particles in the transient object are not satisfied.
Thus the transient object is not treated as a bound aggregate defined in \S \ref{sec:mass}. 
Finally, the hot transient object breaks up into two large aggregates.

The Hill radius $r_\mathrm{H}$ and physical radius $r_\mathrm{c}$ of the largest aggregate are shown in Figure \ref{fig:all}.
The physical radius $r_\mathrm{c}$ is defined as
\begin{equation}
r_\mathrm{c}=\sqrt{\frac{1}{n_{ic}}\sum_{j=1}^{n_{ic}} |\mathbf x_j-\mathbf x_\mathrm{G} |^2},
\end{equation}
where $\mathbf x_j$ is the position of $j$-th particle, $\mathbf x_\mathrm{G}$ is the position of the aggregate's center of mass. The Hill radius $r_\mathrm{H}$ is about 5 times larger than the physical radius $r_\mathrm{c}$. 

In the soft-sphere model (Model 1), the mass of the largest aggregate follows an almost similar evolution.
However, we do not observe the deep cusp observed in the hard-sphere model because there happens to be no impact between the largest and the second-largest aggregate in this simulation.
Note however that the properties of the largest aggregate is sensitive to the random seed in the initial condition.
We observed deep cusps in the plot that corresponds to a collision between the largest and second-largest aggregates in some of the other runs, even with the soft-sphere model.

\subsubsection{Velocity Dispersion \label{sec:veldis}}
The time evolution of the radial velocity dispersion is shown in Figure \ref{fig:all}.
The velocity dispersion decreases monotonically until $t \simeq 1T_\mathrm{K}$ because of inelastic collisions.
When the velocity dispersion becomes sufficiently small ($Q\simeq 2$), gravitational instability occurs and the wake-like structure appears. 
As a result, the gravitational energy is converted to kinetic energy, and the velocity dispersion increases ($t>1T_\mathrm{K}$). 
Once large aggregates are formed, other small field particles are scattered by the large aggregates and the velocity dispersion of field particles becomes large. Finally, the hot thick disk and the large aggregates on the midplane are formed.
In reaction, dynamical friction from field particles decreases the velocity dispersion of the largest aggregate.

\afterpage{\clearpage}  

\subsection{Parameter Dependence}
We summarize how the results are affected by changing parameters: the size of the computational domain $A$, the restitution coefficient $\epsilon$, the optical depth $\tau$, the ratio of Hill radius to the physical diameter $\zeta$, and the duration of collision $T_s$.
We consider 
the ratio of the aggregate mass to the total mass, 
the mass of the largest aggregate, 
the time evolution of the physical radius,
and the radial velocity dispersion of field particles. 

\subsubsection{Size of Computational Domain \label{sec:calcsize}}
\begin{figure}
\plotone{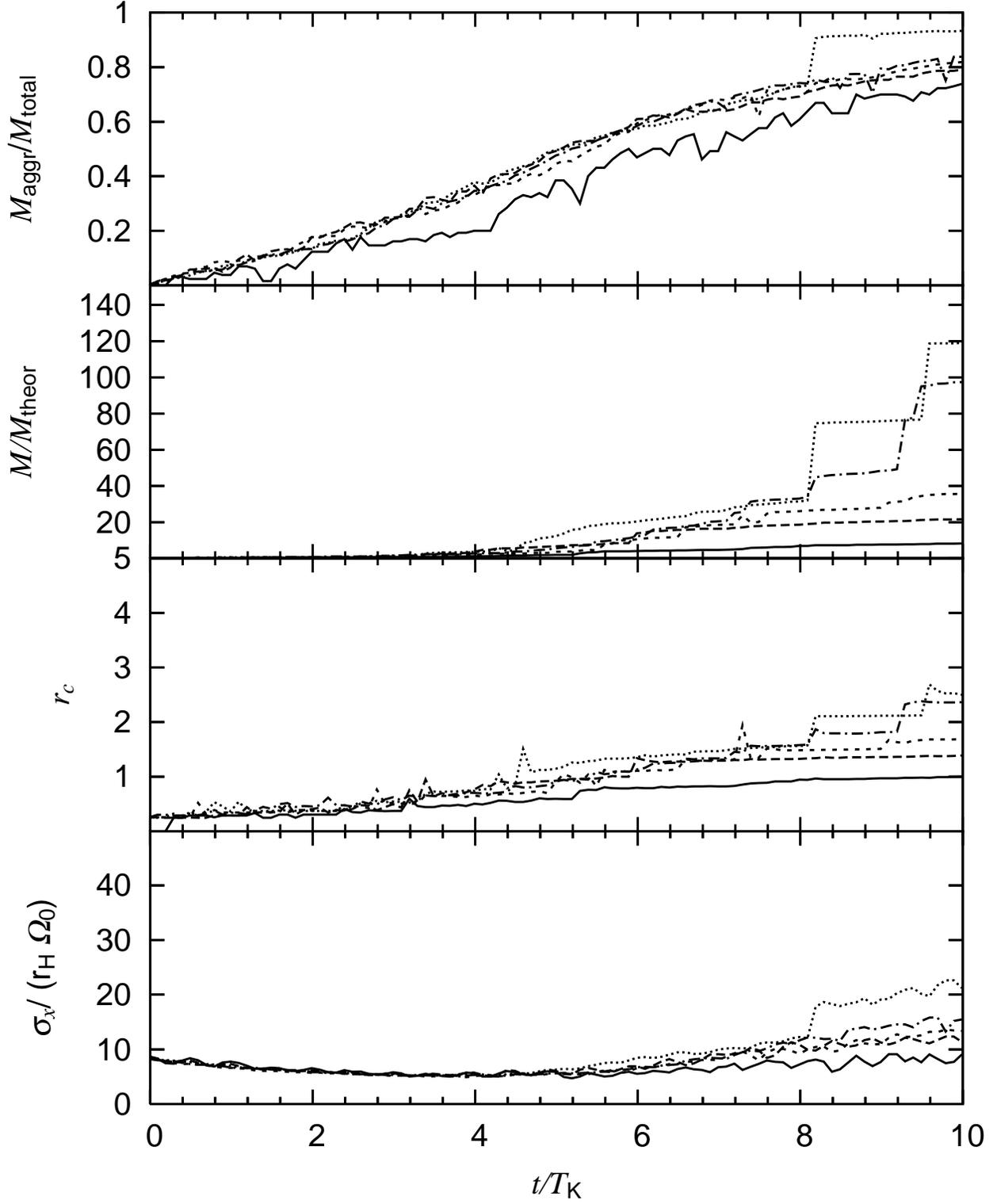}
\caption{The time evolution of 
the ratio of the aggregate mass to the total mass,
the mass of the largest aggregate, 
the physical radius of the largest aggregate,
and the velocity dispersion of field particles
for $A=3$ (\textit{solid curve}), $A=6$ (\textit{dashed curve}), $A=8$ (\textit{short dashed curve}), $A=10$ (\textit{dotted curve}) and $A=12$ (\textit{dash-dotted curve}).}
\label{fig:dep_region_all.eps}
\end{figure}

We set $\tau=0.05$ and  vary the size of computational domain from $A=3$ to $12$ by keeping the other parameters, $\zeta$, and $T_s$ at fiducial value (models 12, 13, 14, 15, 16).
We use shallow optical depth to reduce the number of particles.
We check the dependence of the final mass of the largest aggregate on the size of the computational domain.
As shown in Figure \ref{fig:dep_region_all.eps}, 
the final mass and physical radius of the largest aggregate increase as the size of the computational domain increases.
The ratios of the aggregate mass to the mass for $A=3, 6, 8, 10,$ and $12$ are almost constant, and reach 
about $0.9$ in each model.
The aggregate mass is larger for larger calculation size regions because the total mass increases as the size of the computational domain increases.
The largest aggregate continues to grow until it absorbs most field particles and small aggregates.

There is little difference in the radial velocity dispersion of field particles before GI. 
In contrast, the final radial velocity dispersion of field particles increases with the size of the computational domain.
The scattering of the field particles is more efficient for the larger aggregate.
Therefore, the velocity dispersion increases as the size of the computational domain increases.

We confirmed that the property of the largest aggregate depends on the size of computational domain $L$.
The mass of the largest aggregate increases as the size of computational domain $L$ increases.
The largest aggregate stops growing owing to the depletion of field particles.
Thus, the mass of the largest aggregate in our simulation is not realistic.
We expect that the largest aggregate will possibly stop growing if we perform the simulation on a sufficiently large
computational domain.
We will examine this problem in future work.

\subsubsection{Restitution Coefficient}
The restitution coefficient $\epsilon$ is varied from $\epsilon=0.1$ to $0.9$ (models 7, 8, 9, 10, and 11).
We keep the other parameters, $\tau$, $\zeta$, $A$, and $T_s$ at the fiducial values.
The energy dissipation due to the inelastic collisions becomes inefficient as the restitution coefficient $\epsilon$ increases.
If we ignore the energy gain and consider the dissipation due to inelastic collision only, the damping timescale of the velocity dispersion is described as $T_\mathrm{damp} \simeq c/(\Omega_0 \tau (1 - \epsilon^2))$, where $c$ is a numerical factor of order of unity \citep{Goldreich1978}.
The damping timescales are $T_\mathrm{damp}/T_\mathrm{K} \simeq 1.6, 1.7, 2.1, 3.1, 8.4$ for $\epsilon=0.1, 0.3, 0.5, 0.7,$ and $0.9$, respectively, if we set $c=1$.

\begin{figure}
\plotone{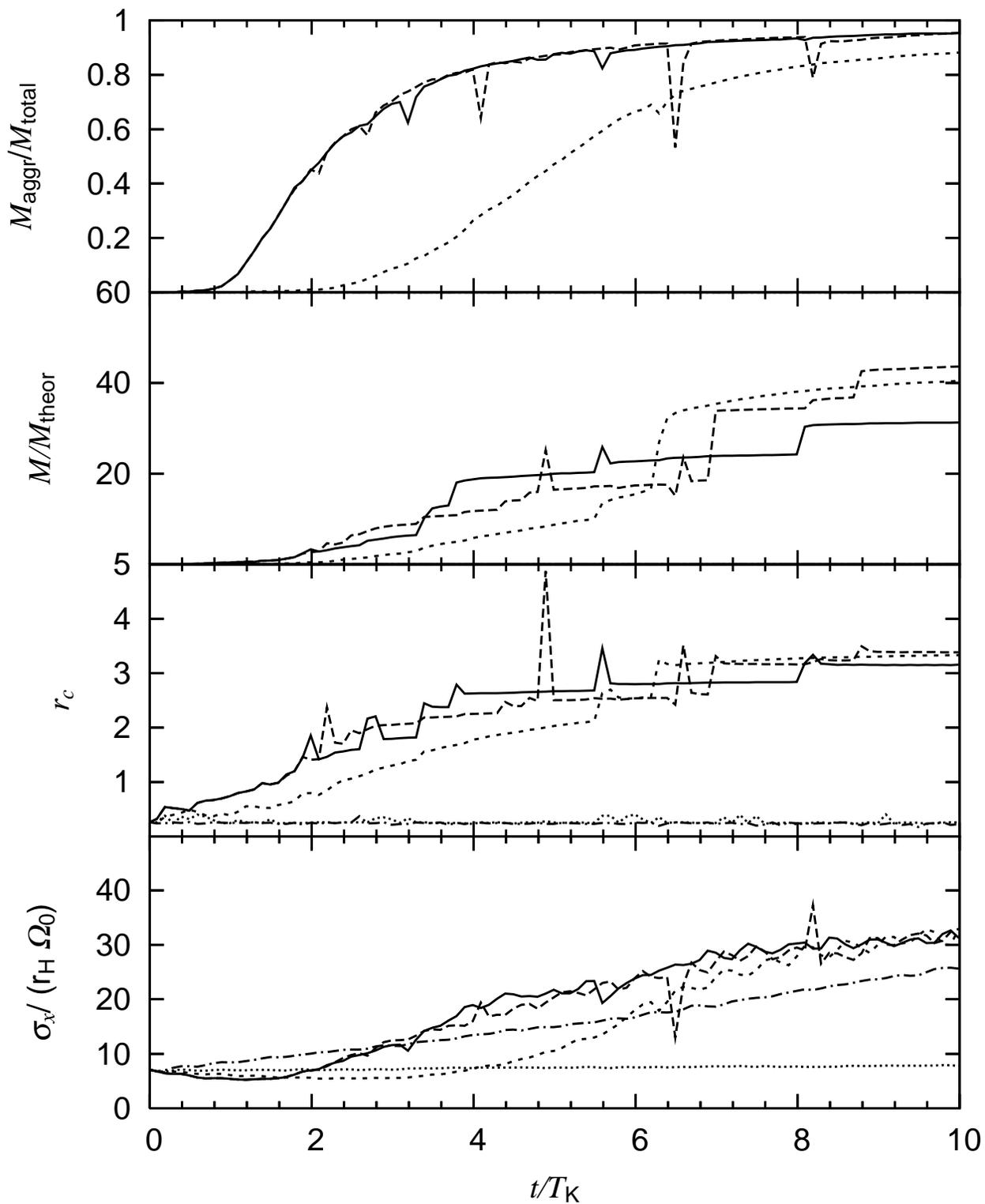}
\caption{The same as Figure \ref{fig:dep_region_all.eps} but for 
$\epsilon=0.1$ (\textit{solid curve}), $\epsilon=0.3$ (\textit{dashed curve}), $\epsilon=0.5$ (\textit{short dashed curve}) , $\epsilon=0.7$ (\textit{dotted dashed curve}), $\epsilon=0.9$ (\textit{dash-dotted curve}).}
\label{fig:dep_epsilon_all.eps}
\end{figure}

As shown in Figure \ref{fig:dep_epsilon_all.eps},
the gravitational instabilities occur only for $\epsilon=0.1, 0.3,$ and $0.5$.
The critical restitution coefficient is $\epsilon_\mathrm{cr} \simeq 0.7$ \citep[e.g.][]{Goldreich1978,Salo1995,Daisaka1999}.
When the restitution coefficient $\epsilon>\epsilon_{cr}$, the dissipation due to collisions is too small to reduce the velocity dispersion. The energy gain from the boundary is more efficient than the dissipation.
In this case, the velocity dispersion increases monotonically, and the velocity dispersion cannot reach the critical value.
Thus no gravitational instability occurs. 

On the other hand, when $\epsilon<\epsilon_c$, the dissipation is efficient and the velocity dispersion can decrease gradually. The decay timescale depends on $\epsilon$ and becomes shorter as $\epsilon$ decreases.
The velocity dispersions become minimum when $T_\mathrm{min}/T_\mathrm{K} \simeq 1.4, 1.4, 2.3$ for $\epsilon=0.1, 0.3, 0.5$, respectively. This dependence of the minimum timescales $T_\mathrm{min}$ on the restitution coefficient agrees approximately with that of the damping timescale $T_\mathrm{damp}$ estimated by the analytic formula. 
When the velocity dispersion reaches the critical value, the gravitational instability occurs.
No remarkable differences are found after the gravitational instability occurs.
The evolution of the size of the largest aggregate, and the ratio of the aggregate mass to the total mass is similar in the models for $\epsilon=0.1, 0.3,$ and $0.5$.

\subsubsection{Optical Depth}
We study the effect of optical depth $\tau$ by comparing results with $\tau=0.05, 0.1,$ and $0.125$ (models 5, 1, and 6). 
The collisions occur frequently for larger $\tau$.
In the case of frequent collisions, the dissipation of kinetic energy due to inelastic collisions is efficient.
The damping timescales of the velocity dispersion of field particles are $T_\mathrm{damp}/T_\mathrm{K}\simeq$ $3.2$, $1.6$, and $1.3$ for $\tau=0.05, 0.1,$ and $0.125$.
The damping timescale becomes long as the optical depth decreases.

\begin{figure}
\plotone{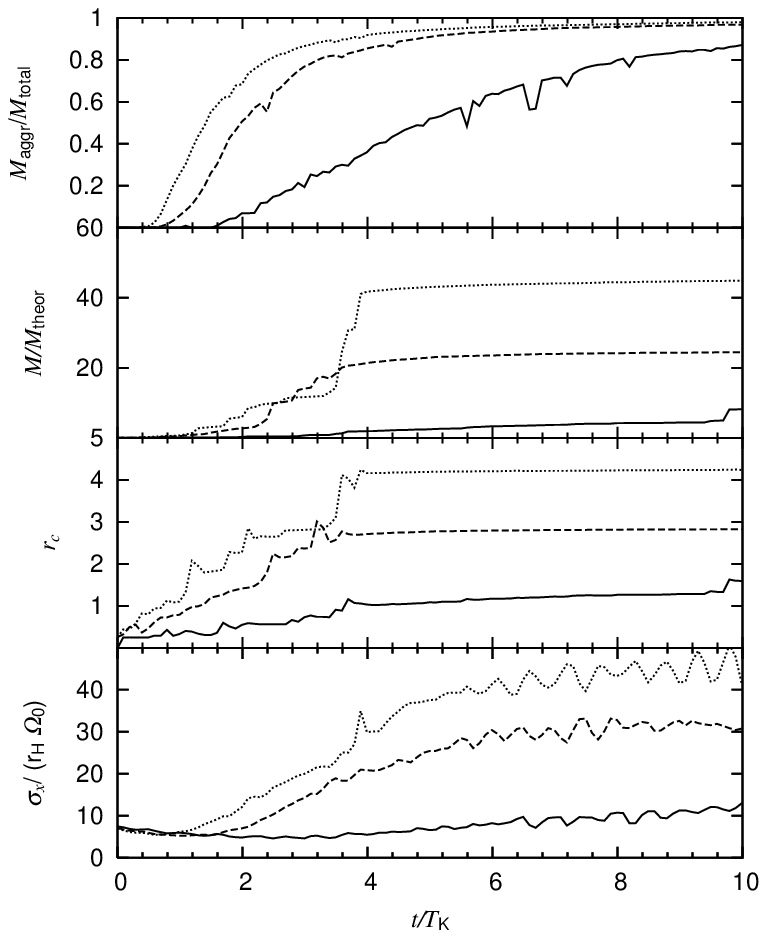}
\caption{The same as Figure \ref{fig:dep_region_all.eps} but for 
$\tau=0.05$ (\textit{solid curve}), $\tau=0.1$ (\textit{dashed curve}), $\tau=0.125$ (\textit{dotted curve}) .}
\label{fig:dep_tau_all.eps}
\end{figure}

As shown in Figure \ref{fig:dep_tau_all.eps}, 
the velocity dispersions of field particles become minimum when $T_\mathrm{min}/T_\mathrm{K} \simeq 3.0, 1.5,$ and $1.0$ for $\tau=0.05, 0.1,$ and  $0.125$, respectively.
This dependence of the timescales $T_\mathrm{min}$ on the optical depth $\tau$ agrees with that of the damping timescale $T_\mathrm{damp}$.
The final velocity dispersion is larger for larger $\tau$. This is because the difference in the mass of the largest aggregates.
The final mass and physical radius of the largest aggregate increases as optical depth $\tau$ increases, as shown in Figure \ref{fig:dep_tau_all.eps}. 
This means that frequent collisions lead to the efficient growth of the largest aggregate.
The ratio of the aggregate mass to the total mass is larger for larger $\tau$. The ratio reaches about $0.9$-$1.0$ for $\tau=0.05$-$1.25$.  
If the collisions occur frequently, the field particles grow rapidly. 

\subsubsection{Hill Radius \label{sec:hr}}
\begin{figure}
\plotone{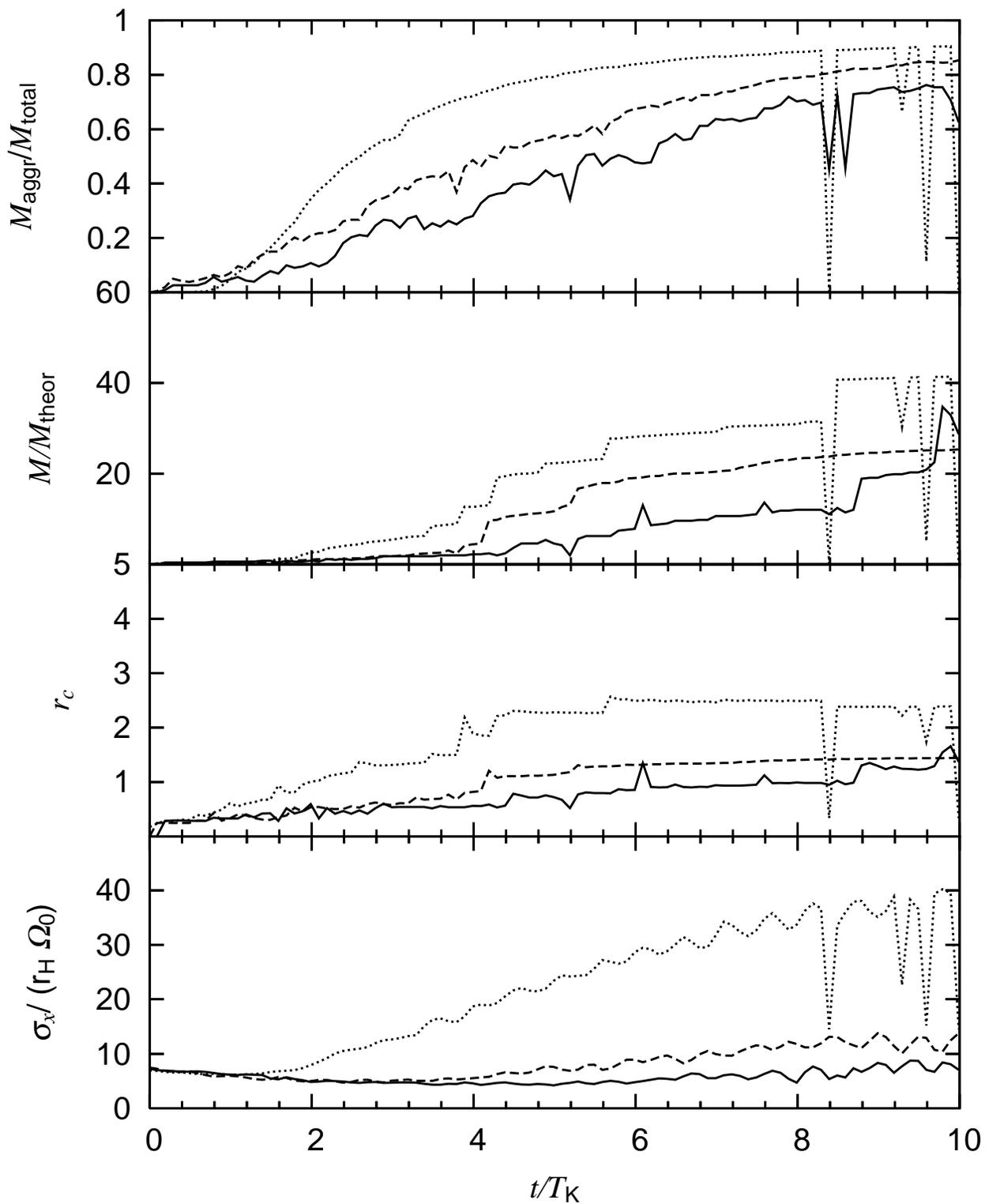}
\caption{The same as Figure \ref{fig:dep_region_all.eps} but for 
$\zeta=1.8$ (\textit{solid curve}), $\zeta=2.0$ (\textit{dashed curve}), $\zeta=3.0$ (\textit{dotted curve}).
}
\label{fig:dep_hill_all.eps}
\end{figure}

We set $\zeta=1.75$, $2.0$, and $3.0$ and fix the other parameters in the fiducial model except for the optical depth $\tau=0.05$ (models 17, 18, and 19). We test the effect of the Hill radius.
The mass and radius of the largest aggregate increase as $\zeta$ increases in Figure \ref{fig:dep_hill_all.eps}.
The Hill radius corresponds to the distance from which the particle effects the other particles through gravitational force.
The number of particles that one particle can affect through mutual gravitational force increases as $\zeta$ increases.
Therefore, the aggregates may absorb field particles more efficiently for larger $\zeta$.
The velocity dispersion of field particles is affected by the large aggregates. 
Therefore, the velocity dispersion of field particles increases as $\zeta$ increases.
The ratio of the aggregate mass to the total mass increases more rapidly for larger $\zeta$. 
From equation (\ref{eq:qq}), the critical velocity dispersion for GI is proportional to $\zeta^2$.
If we start the calculation from the same velocity dispersions, GI occurs earlier for larger $\zeta$.
Therefore, if we set larger $\zeta$, the ratio starts increasing earlier.
The ratios reach about $0.7, 0.8,$ and $0.9$ for $\zeta=1.75, 2.0,$ and $3.0$.
The ratio of the aggregate mass to the total mass is larger if $\zeta$ is larger.

The realistic Hill radius at 1AU corresponds to $\zeta \simeq 100$.
As stated in the above discussion, a larger aggregate may be formed in the case where $\zeta \simeq 100$.
However, in the large $\zeta$ limit, the mass and radius of the largest aggregate must have upper limits
because the mass in the computational domain is fixed.
We will examine the upper limit of the size of the largest aggregate in future work.

\subsubsection{Duration of Collision \label{sec:softhard}}
\begin{figure}
\plotone{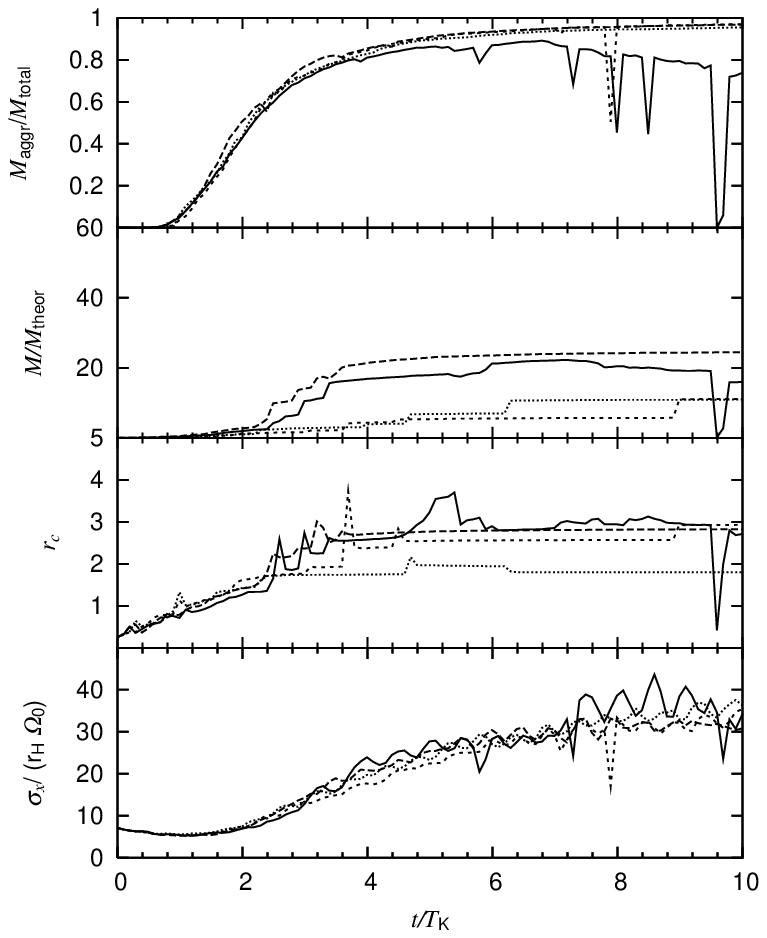}
\caption{The same as Figure \ref{fig:dep_region_all.eps} but for 
the hard-sphere model (\textit{solid curve}), and the soft-sphere model with $T_s=0.01$ (\textit{dashed curve}), $T_s=0.02$ (\textit{short dashed curve}), and $T_s=0.09$ (\textit{dotted curve}).
}
\label{fig:dep_spring_all.eps}
\end{figure}

We compare the hard-sphere and soft-sphere models and vary the duration of collisions for $T_s=0.01, 0.02,$ and $0.09$ (models 1H, 1, 2, and 3) in Figure \ref{fig:dep_spring_all.eps}.
The other parameters are fixed at the fiducial model.
The masses of the largest aggregate reach about $5$-$10 M_\mathrm{theor}$. 
We do not observe any clear dependence of the mass of the largest aggregate on the duration of collision.
This may be caused by the fluctuation.
The time evolutions of velocity dispersion of the field particles are similar for all models. 
There is little difference in the velocity dispersion for the durations of collisions $T_s=0.01, 0.02,$ and $0.09$.
The physical radii for the duration of collision $T_s=0.09$ is smaller than that for the hard-sphere model. 
For the soft-sphere model, the aggregates can contract and reduce their physical radii. 
If the aggregates reduce their radius, the restoring harmonic force of each particle increases.
The balance between the restoring harmonic force and the internal gravitational force of the aggregate determine its physical radius.
The restoring harmonic force decreases as the duration of collision increases.
Therefore the physical radius for the duration of collision $T_s=0.09$ is smaller than that for the other models.

\afterpage{\clearpage}  

\section{Summary\label{sec:sum}}
We performed N-body simulations of the dust layers without a gas component using the sliding cell method.
The gravitational force that has cutoff length is included by using the subregion method.
We adopted the hard-sphere and soft-sphere models as collision models.
We considered the region of the disk around $a_0\simeq 1\mathrm{AU}$ where the dynamics are collision-dominated and the Hill radius of particles to be much larger than their physical size. 
We examined the formation process of planetesimals through the gravitational instability and the time evolution of the fiducial model and its dependence on the system parameters.

If a restitution coefficient $\epsilon$ is smaller than the critical restitution coefficient $\epsilon_\mathrm{cr} \simeq 0.7$, though the velocity dispersion is large, the velocity dispersion decreases owing to inelastic collisions, and gravitational instability occurs when $Q\simeq2$. 
We found that the formation process is divided into three stages: the formation of the non-axisymmetric wake-like structures, the formation of aggregates, and the collisional growth of the aggregates.
The non-axisymmetric wake-like structures are formed due to GI. 
Then the dense parts of wakes fragment into aggregates. 
The aggregates rapidly grow by mutual collisions. Finally, a few large aggregates absorb other small aggregates and field particles.

We found the masses of planetesimals to be larger than these estimated by linear theory.
The mass range of the aggregates is $0.1 M_{\mathrm{theor}} - 1.0 M_{\mathrm{theor}}$ when the GI occurs. 
The mass distribution changes with time because of the rapid growth of aggregates.
The mass of the largest aggregates depends on the size of the computational domain and is larger than $10 M_{\mathrm{theor}}$ in our calculation.

We studied the dependence of the results on the system parameters, the optical depth $\tau$, the ratio of Hill radius to the physical diameter $\zeta$, the size of computational domain $A$, the duration of collision $T_s$, and the restitution coefficient $\epsilon$.
We found that the mass and radius of the largest aggregate increase as $\tau$, $\zeta$, and $A$ increase.
However, the ratio of the aggregate mass to the total mass is $0.8$-$1.0$ and almost independent of them.
The aggregates absorb the most field particles.
The total mass in the size of computational domain increases as $A$ increases.
Therefore the mass of the largest aggregate is larger for larger $A$.
If $\tau$ is large, the collision frequency is large, and accumulation is efficient.
The mutual gravitational force is stronger for larger $\zeta$.
Therefore, the growth of the aggregates is more efficient if $\zeta$ and $\tau$ are larger.
The mass of the largest aggregate is independent of $\epsilon$ and $T_s$.
The physical size of the largest aggregate is smaller for larger $T_s$ since the restoring force is weaker for larger $T_s$.

In order to make the simulation of a self-gravitating dust layer feasible, we had to adopt some simplification and approximation in the present paper.
We used the ``super particle" in the present simulation. 
The super particles represent the group of small dust particles.
We assumed that the super particles obey the simple collision model analogous to that of the small dust particles.
The restitution coefficient $\epsilon$ used in this simulation is not the physical restitution coefficient, but corresponds to the rate of the dissipation due to collisions between super particles. 
Thus, the study of the collision law between super particles and its validity remains to be done.

Unfortunately, the simulation of a self-gravitating dust layer using real-scale dust particles is beyond the ability of today's supercomputers.
We used smaller $\zeta=1.75-3.0$ than the realistic $\zeta \simeq 100$.
We checked that the formation process is common for $\zeta=1.75-3.0$, and investigated its dependence on $\zeta$.
However, it is necessary to test the model with large $\zeta$ to confirm whether our results can be applied to a realistic formation scenario or not.
We will examine this problem in the near future.

We found that the masses of planetesimals increase with the size of
 computational domain in our calculation. 
We did not observe the saturation of this mass increase.
If the saturation does not occur in the actual formation process,
 the protoplanets can be formed directly through GI. 
We need to perform the simulation with a much larger size of the
 computational domain to discuss the saturation of the planetesimal growth.  
  
For the sake of simplicity, we neglected the effect of gas in our
 simulation.  
However, the effect of gas is not negligible if the dust size is
 smaller than about $4$m and is important for the scenario of the
 formation of planetesimals. 
Many authors have investigated the effect of gas:
 the solid particles drift radially due to gas drag
 \citep{Adachi1976,Weidenschilling1977}, angular momentum dissipation
 helps gravitational collapse
 \citep{Ward1976,Youdin2005a,Youdin2005b}, gas turbulence prevents and
 helps concentrating solids
 \citep{Weidenschilling1993,Barge1995,Fromang2006,Johansen2006b},  
 and the clumps of solid particles are formed due to drag instability
 \citep{Youdin2005,Johansen2006b}. 
In addition, the effect of gas must be important for the nonlinear
 process of the gravitational instability. 
If the gas is laminar and the turbulence is sufficiently week, the
 drag force dissipate the kinetic energy of dust particles.  
This effect helps GI. 
On the other hand, if the turbulence of gas is strong, the dust
 particles are stirred up and the kinetic energy increases. 
This effect prevents GI. 
We expect that the criterion and nonlinear process of GI depend on
 the size of the dust particles and the strength of the turbulence. 
We will examine the conditions of gravitational instability and
 formation process of planetesimals in a turbulent gas disk in near
 future. 

\acknowledgments{
This work is supported by the Grant-in-Aid 
(No.15740118, 16077202, 16244120) 
from the Ministry of Education, Culture, Sports, Science, and 
Technology (MEXT) of Japan.
}


\begin{thebibliography}{}
\bibitem[Adachi, Hayashi, \& Nakazawa(1976)]{Adachi1976} Adachi, I., Hayashi, C., \& Nakazawa, K.\ 1976, Progress of Theoretical Physics, 56, 1756 
\bibitem[Balbus \& Hawley(1991)]{Balbus1991} Balbus, S.~A., \& Hawley, J.~F.\ 1991, \apj, 376, 214 
\bibitem[Barge \& Sommeria(1995)]{Barge1995} Barge, P., \& Sommeria, J.\ 1995, \aap, 295, L1 
\bibitem[Brahic(1977)]{Brahic1977} Brahic, A.\ 1977, \aap, 54, 895 
\bibitem[Champney et al.(1995)]{Champney1995}Champney, J. M., A. R. Dobrovolskis, and J. N. Cuzzi, 1995, Phys. Fluids, 7, 1703.
\bibitem[Coradini, Magni, \& Federico(1981)]{Coradini1981} Coradini, A., Magni, G., \& Federico, C.\ 1981, \aap, 98, 173 
\bibitem[Cuzzi et al.(1993)]{Cuzzi1993} Cuzzi, J. N., Dobrovolskis, A. R., \& Champney, J. M. 1993, Icarus, 106, 102.
\bibitem[Daisaka \& Ida(1999)]{Daisaka1999} Daisaka, H., \& Ida, S.\ 1999, Earth, Planets, and Space, 51, 1195 
\bibitem[Daisaka et al.(2001)]{Daisaka2001} Daisaka, H., Tanaka, 
H., \& Ida, S.\ 2001, Icarus, 154, 296 
\bibitem[Dilley(1993)]{Dilley1993} Dilley, J.~P.\ 1993, Icarus, 
105, 225 
\bibitem[Dobrovolskis et al.(1999)]{Dobrovolskis1999}Dobrovolskis, A. R., J. S. Dacles-Mariani, and J. N. Cuzzi, 1999, J. Geophys. Res., 104, 30805
\bibitem[Dubrulle et al.(1995)]{Dubrulle1995} Dubrulle, B., Morfill, G., \& Sterzik, M.\ 1995, Icarus, 114, 237 
\bibitem[Fromang \& Nelson(2006)]{Fromang2006} Fromang, S., \& Nelson, R.~P.\ 2006, \aap, 457, 343 
\bibitem[Furuya(2004)]{Furuya2004} Furuya, I. `Formation of Planetesimals through Gravitational Instability of a Dust Layer ' Kobe University, 2004
\bibitem[Goldreich \& Tremaine(1978)]{Goldreich1978} Goldreich, P., \& Tremaine, S.~D.\ 1978, Icarus, 34, 227 
\bibitem[Goldreich \& Ward(1973)]{Goldreich1973} Goldreich, P.~\& Ward, W.~R.\ 1973, \apj, 183, 1051 
\bibitem[Hayashi(1981)]{Hayashi1981} Hayashi, C.\ 1981, Progress of Theoretical Physics Supplement, 70, 35 
\bibitem[Henon \& Petit(1986)]{Henon1986} Henon, M., \& Petit, J.-M.\ 1986, Celestial Mechanics, 38, 67
\bibitem[Ida \& Makino(1992)]{Ida1992} Ida, S., \& Makino, J.\ 1992, Icarus, 96, 107 
\bibitem[Ishitsu \& Sekiya(2002)]{Ishitsu2002} Ishitsu, N., \& Sekiya, M. 2002, Earth Planets Space, 54, 917
\bibitem[Ishitsu \& Sekiya(2003)]{Ishitsu2003} Ishitsu, N., \& Sekiya, M. 2003, Icarus, 165, 181.
\bibitem[Johansen et al.(2006a)]{Johansen2006a} Johansen, A., Henning, T., \& Klahr, H.\ 2006, \apj, 643, 1219 
\bibitem[Johansen et al.(2006b)]{Johansen2006b} Johansen, A., Klahr, H., \& Henning, T.\ 2006, \apj, 636, 1121
\bibitem[Makino et al.(2003)]{Makino2003} Makino, J., Fukushige, 
T., Koga, M., \& Namura, K.\ 2003, \pasj, 55, 1163 
\bibitem[Michikoshi \& Inutsuka(2006)]{Michikoshi2006} Michikoshi, S., \& Inutsuka, S.\ 2006, \apj, 641, 1131
\bibitem[Nakagawa et al.(1986)]{Nakagawa1986} Nakagawa, Y., Sekiya, M., \& Hayashi, C.\ 1986, Icarus, 67, 375 
\bibitem[Nakazawa \& Ida(1988)]{Nakazawa1988} Nakazawa, K., \& Ida, S.\ 1988, Progress of Theoretical Physics Supplement, 96, 167 
\bibitem[Ohtsuki(1993)]{Ohtsuki1993} Ohtsuki, K.\ 1993, Icarus, 106, 228 
\bibitem[Richardson(1994)]{Richardson1994} Richardson, D.~C.\ 1994, \mnras, 269, 493 
\bibitem[Safronov(1969)]{Safronov1969} Safronov, V. S. 1969, Evolution of the Protoplanetary Cloud and Formation of the Earth and the Planets(Moscow: Nauka)
\bibitem[Salo(1991)]{Salo1991} Salo, H.\ 1991, Icarus, 90, 254 
\bibitem[Salo(1995)]{Salo1995} Salo, H.\ 1995, Icarus, 117, 287 
\bibitem[Sekiya(1983)]{Sekiya1983} Sekiya, M.\ 1983, Progress of Theoretical Physics, 69, 1116 
\bibitem[Sekiya(1998)]{Sekiya1998} Sekiya, M. 1998, Icarus, 133, 298.
\bibitem[Sekiya \& Ishitsu(2000)]{Sekiya2000} Sekiya, M., \& Ishitsu, N. 2000, Earth Planets Space, 52, 517.
\bibitem[Sekiya \& Ishitsu(2001)]{Sekiya2001} Sekiya, M., \& Ishitsu, N. 2001, Earth Planets Space, 53, 761.
\bibitem[Tanga et al.(2004)]{Tanga2004} Tanga, P., Weidenschilling, S.~J., Michel, P., \& Richardson, D.~C.\ 2004, \aap, 427, 1105 
\bibitem[Toomre(1964)]{Toomre1964} Toomre, A.\ 1964, \apj, 139, 1217 
\bibitem[Ward(1976)]{Ward1976} Ward, W.~R. 1976, in {Frontiers of Astrophysics, {\rm ed. E.H. Avrett}},  1--40
\bibitem[Weidenschilling(1977)]{Weidenschilling1977} Weidenschilling, S.~J.\ 1977, \mnras, 180, 57 
\bibitem[Weidenschilling(1980)]{Weidenschilling1980} Weidenschilling, S. J., 1980, Icarus, 44, 172
\bibitem[Weidenschilling \& Cuzzi(1993)]{Weidenschilling1993} Weidenschilling, S.~J., \& Cuzzi, J.~N.\ 1993, Protostars and Planets III, 1031 
\bibitem[Weidenschilling(1995)]{Weidenschilling1995} Weidenschilling, S.~J.\ 1995, Icarus, 116, 433 
\bibitem[Wisdom and Tremaine(1988)]{Wisdom1988} Wisdom, J., \& Tremaine, S.\ 1988, \aj, 95, 925
\bibitem[Yamoto \& Sekiya(2004)]{Yamoto2004} Yamoto, F., \& Sekiya, M.\ 2004, Icarus, 170, 180 
\bibitem[{Youdin}(2005a)]{Youdin2005a} {Youdin}, A.~N. 2005a, \apj, submitted (astro-ph/0508659)
\bibitem[{Youdin}(2005b)]{Youdin2005b} {Youdin}, A.~N. 2005b, \apj, submitted (astro-ph/0508662)
\bibitem[Youdin \& Chiang(2004)]{Youdin2004} Youdin, A.~N.~\& Chiang, E.~I.\ 2004, \apj, 601, 1109 
\bibitem[Youdin \& Goodman(2005)]{Youdin2005} Youdin, A.~N., \& Goodman, J.\ 2005, \apj, 620, 459 
\bibitem[Youdin \& Shu(2002)]{Youdin2002} Youdin, A. N., \& Shu, F. H. 2002, \apj, 580, 494.
\end{thebibliography}
\end{document}